\documentclass[aps,prd,preprint,superscriptaddress,longbibliography,amssymb]{revtex4-1}

\usepackage[utf8]{inputenc}
\usepackage[pdftex]{graphicx}
\usepackage{amsmath}
\usepackage[amssymb,Gray]{SIunits}
\usepackage{textcomp}
\newcommand{\rmi}{\ensuremath{\mathrm{i}}}
\newcommand{\rme}{\ensuremath{\mathrm{e}}}
\newcommand{\abs}[1]{\ensuremath{\left\vert #1 \right\vert}}
\newcommand{\erw}[1]{\ensuremath{\langle #1 \rangle}}

\newcommand{\order}[1]{\mathcal{O}(#1)}
\newcommand{\bleq}{\ensuremath{\mathrel{\phantom{=}}}}
\newcommand{\nnl}{\nonumber\\}
\newcommand{\bra}[1]{\langle #1 \hspace{-2pt} \mid}
\newcommand{\ket}[1]{\mid \hspace{-1pt} #1 \rangle}

\newcommand{\D}{\mathrm{d}}
\renewcommand{\vec}[1]{\mathrm{\mathbf{#1}}}
\newcommand{\up}{\ket{\,\uparrow}}
\newcommand{\down}{\ket{\,\downarrow}}
\newcommand{\kuu}{\ket{\,\uparrow\uparrow}}
\newcommand{\kud}{\ket{\,\uparrow\downarrow}}
\newcommand{\kdu}{\ket{\,\downarrow\uparrow}}
\newcommand{\kdd}{\ket{\,\downarrow\downarrow}}
\newcommand{\buu}{\bra{\uparrow\uparrow\,}}
\newcommand{\bud}{\bra{\uparrow\downarrow\,}}
\newcommand{\bdu}{\bra{\downarrow\uparrow\,}}
\newcommand{\bdd}{\bra{\downarrow\downarrow\,}}
\newcommand{\guu}{\Gamma_{\uparrow\uparrow}}
\newcommand{\gud}{\Gamma_{\uparrow\downarrow}}
\newcommand{\gdu}{\Gamma_{\downarrow\uparrow}}
\newcommand{\gdd}{\Gamma_{\downarrow\downarrow}}
\newcommand{\uu}{{\uparrow\uparrow}}
\newcommand{\ud}{{\uparrow\downarrow}}
\newcommand{\du}{{\downarrow\uparrow}}
\newcommand{\dd}{{\downarrow\downarrow}}
\newcommand{\taua}{\tau_\text{acc}}

\newcommand{\schr}{Schr{\"o}\-din\-ger}

\begin{document}


\title{Acceleration noise constraints on gravity induced entanglement}


%
\author{Andr\'e Gro{\ss}ardt}
\email[]{andre.grossardt@uni-jena.de}
\affiliation{Institute for Theoretical Physics, Friedrich Schiller University Jena, Fr\"obelstieg 1, 07743 Jena, Germany}


\date{\today}

\begin{abstract}
It has been proposed that quantum features of the gravitational field can be exposed experimentally by employing gravity as a mediator of entanglement. We show that in order to witness this type of entanglement experimentally, strong limits on acceleration noise, which has been neglected in previous work, must be overcome. In the case of two particles of similar mass, Casimir-Polder forces lead to a fundamental limit of tenths of $\femto\meter\,\power{s}{-2}/\sqrt{\hertz}$. Limits are between three and six orders of magnitude less strict for two particles of unequal mass, depending on collisional decoherence.
\end{abstract}


\maketitle



\section{Introduction}
Despite tremendous efforts in quantum gravity research, there is no empirical evidence, to date, as to whether or not the gravitational field must be quantized~\cite{rosenfeldQuantizationFields1963,mattinglyQuantumGravityNecessary2005}. Indirect arguments for the necessity of its quantization~\cite{eppleyNecessityQuantizingGravitational1977,pageIndirectEvidenceQuantum1981} are generally considered inconclusive~\cite{mattinglyQuantumGravityNecessary2005,albersMeasurementAnalysisQuantum2008}. Proposals for experimental tests~\cite{carlipQuantumGravityNecessary2008,yangMacroscopicQuantumMechanics2013,grossardtOptomechanicalTestSchrodingerNewton2016} focus on the specific semi-classical model where curvature of a classical spacetime is sourced by the modulus squared of the quantum state~\cite{mollerTheoriesRelativistesGravitation1962,rosenfeldQuantizationFields1963,mattinglyQuantumGravityNecessary2005,bahramiSchrodingerNewtonEquationIts2014}. On the other hand, experiments to test classical gravitational forces in micromechanical systems~\cite{schmoleMicromechanicalProofofprincipleExperiment2016} are still a long way from probing gravitational fields sourced by nonclassical states, leaving a large gap between systems with observed quantum features on the one hand and systems whose gravitational fields have been measured on the other.

Quantum entanglement, which is often considered the most characteristic feature that separates quantum systems from the classical world, may serve as a means to close this gap. For two quantum particles interacting only gravitationally, it is expected that a quantized gravitational field can yield an entangled state, whereas classical spacetime curvature cannot.
In a recent letter, Bose et~al.~\cite{boseSpinEntanglementWitness2017} propose an idea how to use spin as a witness for this type of gravitationally induced entanglement. 
Two spin-\textonehalf\ particles are each put into a spatial superposition state, where one part of the superposition of each particle experiences a gravitational pull depending on the state of the other particle. This results in a conditional phase shift, which can yield nonclassical spin correlations.

As a concrete realization, Bose et~al. propose to use micrometer sized diamonds, initially separated by about \unit{450}{\micro\meter}. In a magnetic field gradient of about $\unit{10^6}{\tesla\per\meter}$ these are split up for about half a second to yield a superposition of about \unit{250}{\micro\meter} each, such that the two closer parts of the superposition approach each other at about \unit{200}{\micro\meter} distance. After moving parallelly for \unit{2.5}{\second}, a reversed magnetic field gradient rejoins both particle states.

These parameters are carefully chosen: distances between particles must remain large enough such that Casimir-Polder forces do not superseed gravitational ones, flight times shorter than the relevant decoherence time scales, and magnetic field gradients technologically feasible, with the gravitational potential still yielding a sufficiently large phase shift.

There is, however, an obvious caveat: as the gravitational acceleration scales with $R^3/L^2$, $R$ being the source mass radius and $L$ its distance, the acceleration resulting from the micrometer particle at \unit{200}{\micro\meter} distance is matched by the gravitational acceleration of a centimeter particle in kilometer distance. Hence, one should ask why an experiment sensitive to the former should not be influenced by the latter.

As long as both the particles and the experimental set-up, including the magnetic field gradient, are in perfect free fall, the equivalence principle prevents any observable effect of external homogeneous gravitational fields. The proposed experiment, therefore, is ideally performed in a zero gravity environment. If, however, external forces act on the particles and the rest of the experiment differently, such that either the particles or the magnetic fields experience an acceleration relative to the geodesic motion of the center of gravity of the entire experiment, one ends up with a residual observable phase. This is essentially the famous COW experiment~\cite{colellaObservationGravitationallyInduced1975} (althoug in a much weaker gravitational potential), where a gravitational phase can be interpreted either as the consequence of the Newtonian gravitational potential in the laboratory frame or as the effect of accelerated mirrors in the co-moving frame~\cite{greenbergerNeutronInterferometerDevice1983}.

Residual acceleration cannot be entirely avoided even in zero gravity experiments, where it can be expressed in the form of noise spectra that for sub-Hertz frequencies resemble white noise. For drop tower experiments on Earth this residual acceleration reaches typical values around $\unit{10^{-5}}{\meter\per\second\squared}$ (hence ``micro''gravity). Selig et~al.~\cite{seligDropTowerMicrogravity2010} describe improvements in preparation for the MICROSCOPE space mission for testing the weak equivalence principle, finding an approximately frequency-independent noise spectrum around $\sqrt{S_0} \sim \unit{10^{-7}}{\meter\,\power{\second}{-2}/\sqrt{\hertz}}$ for frequencies below \unit{10}{\hertz}.
As far as experiments in space are concerned, the LISA pathfinder mission~\cite{armanoSubfemtogFreeFall2016} minimized acceleration noise as one of their main objectives. The acceleration noise spectrum shows a frequency-independent value around $\sqrt{S_0} \sim \unit{5.6}{\femto\meter\,\power{\second}{-2}/\sqrt{\hertz}}$ in the sub-Hertz range, with a significant increase for frequencies above \unit{0.1}{\hertz}. We will use these two values as a reference for feasible noise levels on Earth and in space, respectively.

We derive the relevant phases contributing to the wave function in section~\ref{sec:grav-phase}, including the decohering effect of noise from external acceleration in subsection~\ref{sec:accel}. In section~\ref{sec:constraints} we give constraints that result from this acceleration noise contribution, first for the case of two similar masses where the distance of closest approach determined by the Casimir-Polder interaction poses a fundamental limit of $\unit{0.24}{\femto\meter\,\power{s}{-2}/\sqrt{\hertz}}$ on the acceptable acceleration noise, then for the case of different masses, where we show that for the parameters chosen by Bose et~al.~\cite{boseSpinEntanglementWitness2017} acceleration noise must still remain below $\unit{1.4}{\pico\meter\,\power{\second}{-2}/\sqrt{\hertz}}$. Finally, we discuss the consequences of these results and possible loopholes in section~\ref{sec:discussion}.

\section{Gravitational phase in spatial superpositions}\label{sec:grav-phase}
It is well known, not least from the COW experiment~\cite{colellaObservationGravitationallyInduced1975}, that a particle in superposition at different levels of a gravitational potential experiences a phase shift. This phase shift can be derived as a perturbative effect around the quasi-classical trajectory.

As in the experimental scenario envisioned by Bose et al.~\cite{boseSpinEntanglementWitness2017}, we consider two spin-\textonehalf\ particles at positions $\vec r(t)$ and $\vec s(t)$, respectively, which for the remainder of this article will be also labelled by $r$ and $s$. The initial state
\begin{equation}
 \ket{\Psi}_0 = \frac{1}{2}\left(\kuu+\kud+\kdu+\kdd\right) \otimes \ket{\vec r_0}\otimes \ket{\vec s_0}
\end{equation}
is subject to a magnetic field gradient, entangling spin and position:
\begin{equation}\label{eqn:general-form-of-state}
\ket{\Psi}_t \sim a\kuu \ket{\vec r_\uparrow}_t \ket{\vec s_\uparrow}_t
+ b\kud \ket{\vec r_\uparrow}_t \ket{\vec s_\downarrow}_t
+ c\kdu \ket{\vec r_\downarrow}_t \ket{\vec s_\uparrow}_t
+ d\kdd \ket{\vec r_\downarrow}_t \ket{\vec s_\downarrow}_t\,.
\end{equation}
The states $\ket{\vec r_\uparrow}_t$ and so forth refer to the evolution of the center of mass for the respective spin and are assumed to be well focused around the positions $\vec r_\uparrow$ etc.
In an accelerated frame with time dependent acceleration $\vec g(t)$, the total state $\ket{\Psi}$ evolves according to the \schr\ equation
\begin{subequations}\label{eqn:schroedinger}\begin{align}
 \rmi \hbar \partial_t \ket{\Psi}_t &= (\hat{H}_0 + \hat{\Gamma}_t) \ket{\Psi}_t \\
 \hat{H}_0 &= \frac{\hat{\vec p}_r^2}{2 m_r} + \frac{\hat{\vec p}_s^2}{2 m_s}
 + \mu_B \, \sigma(t) \, \partial_x B \,(\hat{\sigma}^{r}_x \,\hat{r}_x + \hat{\sigma}^{s}_x\hat{s}_x)\\
 \hat{\Gamma}_t &= -\frac{G m_r m_s}{\abs{\hat{\vec r}-\hat{\vec s}}} - \vec g(t) \cdot (m_r \hat{\vec{r}} + m_s \hat{\vec{s}}) \,.
\end{align}\end{subequations}
$\hat{H}_0$ is the gravitation free Hamiltonian containing both the inertial evolution and the inhomogeneous magnetic field whose action can be flipped or turned off via spin flips or transitions from electron to nuclear spin described by $\sigma(t) \in \{-1,0,1\}$. $\hat{\Gamma}_t$ contains the internal gravitational interaction, as well as the external acceleration $\vec g(t)$.

Note that, against usual convention, $\up$ and $\down$ in $\kuu = \up^{r} \otimes \up^{s}$, and so on, denote $\hat{\sigma}_x$ eigenstates. As $\hat{\Gamma}_t$ does not act on the spin, and the states $\ket{\vec r_\uparrow}_t$, and so on, are approximately position eigenstates, the four parts of the state~\eqref{eqn:general-form-of-state} are approximate eigenstates of $\hat{\Gamma}_t$. Hence we can express the action of the operator $\hat{\Gamma}_t$ by its (time dependent) eigenvalues
\begin{equation}
\guu = -\frac{G m_r m_s}{\abs{\vec r_\uparrow-\vec s_\uparrow}} - \vec g(t) \cdot (m_r \vec r_\uparrow + m_s \vec s_\uparrow)\,,
\end{equation}
and accordingly for $\gud$, $\gdu$, and $\gdd$.
For the subsequent discussion, we absorb the position states $\ket{\vec r_\uparrow}_t \ket{\vec s_\uparrow}_t$ in the spin states, writing $\kuu = \kuu \ket{\vec r_\uparrow}_t \ket{\vec s_\uparrow}_t$, and so forth.

If the potential $\hat{\Gamma}_t$ is small compared to the kinetic energy of the particles, then the classical trajectories $\ket{\vec r_\uparrow}_t$, $\ket{\vec r_\downarrow}_t$, $\ket{\vec s_\uparrow}_t$, and $\ket{\vec s_\downarrow}_t$ are the solution of the unperturbed \schr\ equation $\rmi \hbar \partial_t \ket{\Psi_0}_t = \hat{H}_0 \ket{\Psi_0}_t$, and the perturbation $\hat{\Gamma}_t$ only yields a phase:
\begin{subequations}\begin{align}
 \ket{\Psi}_t &= \frac{1}{2} \left(\rme^{\rmi \phi_{\uu}} \kuu + \rme^{\rmi \phi_{\ud}} \kud + \rme^{\rmi \phi_{\du}} \kdu + \rme^{\rmi \phi_{\dd}} \kdd \right) \\
\phi_{\uu}
&= \frac{G m_r m_s}{\hbar} \int_0^t \frac{\D t'}{\abs{\vec r_\uparrow(t') - \vec s_\uparrow(t')}}
+ \frac{1}{\hbar} \int_0^t \D t' \, \vec g(t') \cdot (m_r \vec r_\uparrow(t') + m_s \vec s_\uparrow(t')) \,,
\end{align}\end{subequations}
and accordingly for $\phi_{\ud}$, $\phi_{\du}$, and $\phi_{\dd}$.

\subsection{Particle trajectories}
We choose as a reference frame the initial rest frame of the two particles, with the $x$-axis defined by the particle positions at time $t=0$:
\begin{equation}
 \vec r_\uparrow(0) = \vec r_\downarrow(0) = \vec r_0 = (-d/2,0,0) \quad\quad\text{and}\quad\quad
 \vec s_\uparrow(0) = \vec s_\downarrow(0) = \vec s_0 = (d/2,0,0) \,.
\end{equation}
Let the inhomogeneous magnetic field used to create the spatial superposition be aligned with this reference frame, resulting in an acceleration
\begin{equation}
 \vec a^{r,s}_\uparrow(t) = -\frac{2\Delta x_{r,s}}{\taua^2} \sigma(t) \vec e_x \quad\quad\text{and}\quad\quad
 \vec a^{r,s}_\downarrow(t) = \frac{2\Delta x_{r,s}}{\taua^2} \sigma(t) \vec e_x\,,
\end{equation}
with
\begin{equation}\label{eqn:delta-x-dep-field-gradient}
\Delta x_{r,s} = \frac{\mu_B\,\partial_x B \,\taua^2}{2 m_{r,s}} \,,
\end{equation}
where $\vec e_x$ is the unit vector in $x$-direction, $\mu_B$ the Bohr magneton, and
\begin{equation}
 \sigma(t) = \begin{cases} 1 & \text{for } t \in [0,\frac{\taua}{2}] \cup [\tau+\frac{3\taua}{2},\tau+2\taua] \\
 -1 &  \text{for } t \in [\frac{\taua}{2},\taua] \cup [\tau+\taua,\tau+\frac{3\taua}{2}] \\
 0 & \text{everywhere else.} \end{cases}
\end{equation}
The trajectories of the four parts of the wave function, defined by $\hat{H}_0$, are
\begin{equation}
 \vec r_\uparrow(t) = \vec r_0 + \int_0^t \D t' \, \int_0^{t'} \D t'' \, \vec a^r_\uparrow(t'') \,,
\end{equation}
and accordingly for the other three trajectories. For the particle distances we find:
\begin{subequations}\label{eqn:distances}\begin{align}
\abs{\vec r_\uparrow(t) - \vec s_\uparrow(t)} &= \abs{\vec r_0 - \vec s_0 
+ \int_0^t \D t' \, \int_0^{t'} \D t'' \, (\vec a^r_\uparrow(t'') - \vec a^s_\uparrow(t'') )} \nnl
&= d + \frac{1}{2}(\Delta x_r - \Delta x_s) \Sigma_\text{acc}(t)
= d +  \delta x \Sigma_\text{acc}(t)\\
\abs{\vec r_\uparrow(t) - \vec s_\downarrow(t)} &= { d + \frac{1}{2}(\Delta x_r + \Delta x_s) \Sigma_\text{acc}(t) }
= d +  \Delta x \Sigma_\text{acc}(t) \\
\abs{\vec r_\downarrow(t) - \vec s_\uparrow(t)} &= { d - \frac{1}{2}(\Delta x_r + \Delta x_s) \Sigma_\text{acc}(t) } 
= d -  \Delta x \Sigma_\text{acc}(t) \\
\abs{\vec r_\downarrow(t) - \vec s_\downarrow(t)} &= { d - \frac{1}{2}(\Delta x_r - \Delta x_s) \Sigma_\text{acc}(t) }
= d -  \delta x \Sigma_\text{acc}(t) \,,
\end{align}\end{subequations}
where we introduce $\Delta x=(\Delta x_r + \Delta x_s)/2$, $\delta x =(\Delta x_r - \Delta x_s)/2$, as well as
\begin{equation}
\Sigma_\text{acc}(t) = \frac{4}{\taua^2} \int_0^t \D t' \, \int_0^{t'} \D t'' \, \sigma(t'')
= \begin{cases}
\frac{2t^2}{\taua^2} & \text{for } t \in [0, \frac{\taua}{2}] \\
\frac{4t}{\taua}-\frac{2t^2}{\taua^2} - 1 & \text{for } t \in [\frac{\taua}{2} , \taua] \\
1 & \text{for } t \in [\taua , \tau + \taua] \\
 - \frac{2(t-\tau)(t-\tau-2\taua)}{\taua^2}-1& \text{for } t \in [\tau + \taua , \tau + \frac{3\taua}{2}] \\
\frac{2(t-\tau-2\taua)^2}{\taua^2} & \text{for } t \in [\tau + \frac{3\taua}{2} , \tau + 2 \taua] \,.
  \end{cases}
\end{equation}
Note that $\Sigma_\text{acc}(\tau+2\taua-t) = \Sigma_\text{acc}(t)$ and the absolute values in equations~\eqref{eqn:distances} can be omitted since $\Delta x < d$, $\delta x < d$, and $0 \leq \Sigma_\text{acc}(t) \leq 1$.

\subsection{Associated phases}
We split up each of the phases into the contributions from the mutual gravitational interaction during the free flight time $\tau$, during the initial and final acceleration periods, as well as the phases for each trajectory due to the external acceleration:
\begin{subequations}\label{eqn:phases-complete}\begin{align}
\phi_\uu &= \phi^\tau_\uu + \phi^\text{acc}_\uu + \phi^\text{ext}_{r\uparrow} + \phi^\text{ext}_{s\uparrow} \\
\phi_\ud &= \phi^\tau_\ud + \phi^\text{acc}_\ud + \phi^\text{ext}_{r\uparrow} + \phi^\text{ext}_{s\downarrow} \\
\phi_\du &= \phi^\tau_\du + \phi^\text{acc}_\du + \phi^\text{ext}_{r\downarrow} + \phi^\text{ext}_{s\uparrow} \\
\phi_\dd &= \phi^\tau_\dd + \phi^\text{acc}_\dd + \phi^\text{ext}_{r\downarrow} + \phi^\text{ext}_{s\downarrow} \,.
\end{align}\end{subequations}
The phases collected during the free flight are
\begin{subequations}\begin{align}
\phi^\tau_\uu &= \frac{G m_r m_s}{\hbar} \int_{\taua}^{\tau+\taua} \frac{\D t}{{ d + \frac{1}{2}(\Delta x_r - \Delta x_s)} }
= \frac{G m_r m_s \tau}{\hbar (d + \delta x)} \\
\phi^\tau_\ud &= \frac{G m_r m_s \tau}{\hbar (d + \Delta x)} \\
\phi^\tau_\du &= \frac{G m_r m_s \tau}{\hbar (d - \Delta x)} \\
\phi^\tau_\dd &= \frac{G m_r m_s \tau}{\hbar (d - \delta x)} \,.
\end{align}\end{subequations}
For the phase due to external acceleration we find
\begin{subequations}\begin{align}
\phi^\text{ext}_{r\uparrow}(t) &=
\frac{m_r}{\hbar} \int_0^t \D t' \, \vec g(t') \cdot \vec r_\uparrow(t') \nnl
&=
 \frac{m_r}{\hbar} \int_0^t \D t' \, \vec g(t') \cdot \vec r_0 
 + \frac{m_r}{\hbar} \int_0^t \D t' \, \int_0^{t'} \D t'' \, \int_0^{t''} \D t''' \, \vec g(t') \cdot \vec a^r_\uparrow(t''') \nnl
&= -\frac{m_r d}{2 \hbar} v_x(t) + \frac{1}{2}\chi(t) \\
\phi^\text{ext}_{r\downarrow}(t) &= -\frac{m_r d}{2\hbar}  v_x(t) - \frac{1}{2}\chi(t)  \\
\phi^\text{ext}_{s\uparrow}(t) &= \frac{m_s d}{2\hbar} v_x(t) + \frac{1}{2}\chi(t)  \\
\phi^\text{ext}_{s\downarrow}(t) &= \frac{m_s d}{2\hbar} v_x(t) - \frac{1}{2}\chi(t)  \,,
\end{align}\end{subequations}
with the velocity
\begin{align}
 v_x(t) &= \int_0^t \D t' \, g_x(t') \,, \quad\text{as well as} \\
 \chi(t) &= \frac{2\,\mu_B\,\partial_x B}{\hbar} \int_0^t \D t' \, \int_0^{t'} \D t'' \, \int_0^{t''} \D t''' \, g_x(t') \sigma(t''') \label{eqn:phase-chi} \,.
\end{align}
We calculate the phases collected during the acceleration period in the appendix.

\subsection{Effect of random external acceleration}\label{sec:accel}

Let us now address the phase $\chi$ due to the external acceleration. We can simplify equation~\eqref{eqn:phase-chi}
\begin{align}\label{eqn:phase-chi-simplified}
 \chi &= \frac{\mu_B\,\partial_x B \, \taua^2}{2\hbar} \int_0^{\tau+2\taua} \D t \, g_x(t)  \Sigma_\text{acc}(t) \nnl
 &= \frac{\mu_B\,\partial_x B \, \taua^2}{\hbar} \int_0^{\taua} \D t \, g_x(t)  \Sigma_\text{acc}(t)
 + \frac{\mu_B\,\partial_x B \, \taua^2}{2 \hbar} \int_{\taua}^{\tau+\taua} \D t \, g_x(t) \,,
\end{align}
and for $\taua \ll \tau$ we can neglect the first part due to the acceleration phase. Evidently, only the component $g_x$ parallel to the $x$-axis (defined by the initial particle positions) affects the phase shift. The experimental set-up will generally be chosen in such a way that the time average is $\erw{g_x(t)} \approx 0$, for instance by aligning the field gradient and the particles parallel to the surface of the Earth. However, there will be fluctuations of $g_x$ in time which can be associated with a noise spectrum $S(\omega)$ through the correlation functions
\begin{equation}
 \erw{g_x(0) g_x(t)} = \int \frac{\D\omega}{2\pi} S(\omega) \rme^{-\rmi \omega t} \,.
\end{equation}
We assume that the acceleration noise is well approximated by white Gaussian noise, i.e.\ $S(\omega) \approx S_0$. To obtain the variance of the phase $\chi$ over the averaging time $\tau$, according to equation~\eqref{eqn:phase-chi-simplified}, one can apply a low-pass filter with bandwidth $1/\tau$~\cite{lamineGravitationalDecoherenceAtomic2002} yielding
\begin{equation}
\Delta \chi^2 = \left(\frac{\mu_B\,\partial_x B \, \taua^2}{2 \hbar}\right)^2 \tau^2 \int \frac{\D\omega}{2\pi}  \frac{S_0}{1 + \omega^2 \tau^2}
= \left(\frac{\mu_B\,\partial_x B \, \taua^2}{4 \hbar}\right)^2 S_0 \tau \,.
\end{equation}
We find that in repeated measurements the phase will be distributed around $\chi = 0$ with a probability density
\begin{equation}\label{eqn:pdf-chi}
P(\chi) = \frac{\exp\left(-\frac{\chi^2}{2 \Delta \chi^2}\right)}{\sqrt{2 \pi \Delta \chi^2}} \,.
\end{equation}
Instead of the state~\eqref{eqn:final-state}, we will end up with a classical mixture described by a density matrix
\begin{equation}\label{eqn:density-matrix}
 \rho = \int \D \chi \,P(\chi)  \ket{\Psi(\chi)} \bra{\Psi(\chi)} \,.
\end{equation}

\section{Constraints on witnessing entanglement}\label{sec:constraints}
Let us now first focus on the situation where both particles have similar masses, $m_r \approx m_s$, as proposed by Bose et~al.~\cite{boseSpinEntanglementWitness2017}. In this case, we also find $\Delta x \approx \Delta x_r \approx \Delta x_s$ and $\delta x \ll \Delta x < d$.
If we extract from the phases~\eqref{eqn:phases-complete} the global phase
\begin{equation}\label{eqn:global-phase-similar-masses}
\phi = \frac{G m_r m_s \tau}{\hbar d} + \frac{2 G m_r m_s \taua}{\hbar d} + \frac{m_s-m_r}{\hbar} d \, v_x  
\end{equation}
we can write the final state as
\begin{equation}\label{eqn:final-state}
 \ket{\Psi}_\tau = \frac{\rme^{\rmi \phi}}{2} \left(\rme^{\rmi \widetilde{\chi}} \kuu 
 + \rme^{\rmi \delta\phi} \kud 
 + \rme^{\rmi \Delta\phi} \kdu 
 + \rme^{-\rmi \widetilde{\chi}} \kdd \right) \,.
\end{equation}
with
\begin{subequations}\label{eqn:phases-similar-mass}\begin{align}
\Delta \phi &= \frac{G m_r m_s \tau}{\hbar (d-\Delta x)} - \frac{G m_r m_s \tau}{\hbar d} + \Delta\phi^\text{acc}_-\\
\delta \phi &= \frac{G m_r m_s \tau}{\hbar (d+\Delta x)} - \frac{G m_r m_s \tau}{\hbar d} -\Delta\phi^\text{acc}_+\\
\widetilde{\chi} &= \chi - \delta\chi \,.
\end{align}\end{subequations}
The phases $\Delta\phi^\text{acc}_\pm$ and $\delta \chi$ are given in the appendix.

\subsection{Entanglement witness}
From the probability density function~\eqref{eqn:pdf-chi}, making use of the relation
\begin{equation}
\int_{-\infty}^\infty P(\chi) \rme^{\rmi k \chi} \D \chi = \rme^{-\frac{k^2}{2} \Delta \chi^2}
= \rme^{- k^2 \gamma} \quad\quad\text{with}\quad\quad
\gamma = \frac{\Delta \chi^2}{2} \approx \frac{m_r m_s \, \Delta x^2 \, \tau \, S_0}{8 \, \hbar^2}\,,
\end{equation}
we find the density matrix
\begin{align}\label{eqn:density}
 \rho &= \int \D \chi \,P(\chi)  \ket{\Psi(\chi)} \bra{\Psi(\chi)}  \nnl
 &= \frac{1}{4} \int \D \chi \,P(\chi) \Big[
 \rme^{\rmi \widetilde{\chi}} \kuu \left(\rme^{-\rmi \widetilde{\chi}} \buu 
 + \rme^{-\rmi \delta\phi} \bud 
 + \rme^{-\rmi \Delta\phi} \bdu 
 + \rme^{\rmi \widetilde{\chi}} \bdd \right) \nnl
 &\bleq + \rme^{\rmi \delta\phi} \kud \left(\rme^{-\rmi \widetilde{\chi}} \buu 
 + \rme^{-\rmi \delta\phi} \bud 
 + \rme^{-\rmi \Delta\phi} \bdu 
 + \rme^{\rmi \widetilde{\chi}} \bdd \right) \nnl
 &\bleq + \rme^{\rmi \Delta\phi} \kdu \left(\rme^{-\rmi \widetilde{\chi}} \buu 
 + \rme^{-\rmi \delta\phi} \bud 
 + \rme^{-\rmi \Delta\phi} \bdu 
 + \rme^{\rmi \widetilde{\chi}} \bdd \right) \nnl
 &\bleq + \rme^{-\rmi \widetilde{\chi}} \kdd \left(\rme^{-\rmi \widetilde{\chi}} \buu 
 + \rme^{-\rmi \delta\phi} \bud 
 + \rme^{-\rmi \Delta\phi} \bdu 
 + \rme^{\rmi \widetilde{\chi}} \bdd \right) \Big] \nnl
&= \frac{1}{4} \left(\begin{array}{cccc}
1 & \rme^{-\gamma - \rmi (\delta \phi + \delta\chi)} & \rme^{-\gamma - \rmi (\Delta \phi+\delta\chi)} & \rme^{-4\gamma-2\rmi \delta\chi} \\
\rme^{-\gamma + \rmi (\delta \phi+\delta\chi)} & 1 & \rme^{-\rmi (\Delta \phi - \delta \phi)} & \rme^{-\gamma + \rmi (\delta \phi-\delta\chi)} \\
\rme^{-\gamma + \rmi (\Delta \phi+\delta\chi)} & \rme^{\rmi (\Delta \phi - \delta \phi)} & 1 & \rme^{-\gamma + \rmi (\Delta \phi-\delta\chi)} \\
\rme^{-4\gamma+2\rmi \delta\chi} & \rme^{-\gamma - \rmi (\delta \phi-\delta\chi)} & \rme^{-\gamma - \rmi (\Delta \phi-\delta\chi)} & 1
\end{array}\right) 
\end{align}
in the basis $\{\kuu,\kud,\kdu,\kdd\}$. With this density matrix, we calculate the expectation values
\begin{subequations}\label{eqn:expval}\begin{align} \label{eqn:expval-xz}
\erw{\sigma_x \otimes \sigma_z}_\tau &= \frac{1}{2} \rme^{-\gamma} \left(\cos (\Delta \phi+\delta\chi) - \cos (\delta \phi-\delta\chi)\right)\\
\label{eqn:expval-yy}
\erw{\sigma_y \otimes \sigma_y}_\tau &= \frac{1}{2} \left( \cos(\Delta \phi - \delta \phi) - \rme^{-4 \gamma}\cos(2\delta\chi) \right) \,.
\end{align}\end{subequations}
We define the same entanglement witness $\mathcal{W}$ as in reference~\cite{boseSpinEntanglementWitness2017}, for which
\begin{equation}
 \mathcal{W} = \abs{\erw{\sigma_x \otimes \sigma_z}_\tau + \erw{\sigma_y \otimes \sigma_y}_\tau}
 \leq \frac{1}{2} \left(1 + 2 \rme^{-\gamma} + \rme^{-4 \gamma}\right) \,.
\end{equation}
Evidently, for $\gamma = 0$ one recovers the result by Bose et~al., that $0 \leq \mathcal{W} \leq 2$. However, for finite $\gamma$, in order to find $\mathcal{W} > 1$ and, therefore, evidence for nonclassical behavior, one requires $\gamma \lesssim 0.75$ or
\begin{equation}\label{eqn:s0-limit}
 S_0 \lesssim \frac{8\, \gamma_\text{max}\, \hbar^2}{m_r m_s \tau \, \Delta x^2} \approx \frac{6 \hbar^2}{m_r m_s \tau \, \Delta x^2} \,.
\end{equation}

\subsection{Approximations for the gravitational phases}
Equation~\eqref{eqn:s0-limit} puts a limit on witnessing entanglement. To stay below the limit, where acceleration noise constraints the entanglement witness to values below unity, either the masses, or flight time $\tau$, or the superposition size $\Delta x$ must be sufficiently small. On the other hand, those exact parameters need to be sufficiently large for the gravitational phase to be significantly different from zero.

In order to give a quantitative estimate, we write $a = d-\Delta x$ for the minimal distance between the states $\down^r$ and $\up^s$, and we distinguish three different parameter regimes:

\paragraph{Small superpositions, $\Delta x \ll d$.}
This implies $a \approx d$, therefore $\Delta x \ll a$. Including terms up to quadratic order in $\Delta x/a$, together with the acceleration phases that can be found in the appendix, from equations~\eqref{eqn:phases-similar-mass} we have the phases
\begin{subequations}\label{eqn:phases-small}\begin{align}
\Delta \phi &= \frac{G m_r m_s \tau}{\hbar a} \, \left[\left(1+\frac{\taua}{\tau}\right) \, \frac{\Delta x}{a} - \left(1-\frac{23\,\taua}{30\,\tau}\right) \, \left(\frac{\Delta x}{a}\right)^2 \right] \\
\delta \phi &= -\frac{G m_r m_s \tau}{\hbar a} \, \left[\left(1+\frac{\taua}{\tau}\right) \, \frac{\Delta x}{a} - \left(3+\frac{23\,\taua}{30\,\tau}\right) \, \left(\frac{\Delta x}{a}\right)^2 \right]\,.
\end{align}\end{subequations}
To linear order, we have $\Delta \phi = -\delta \phi$ which implies $\erw{\sigma_x \otimes \sigma_z}_\tau = 0$ and, therefore, $\mathcal{W} \leq 1$. In fact, even when including higher order terms one finds $\mathcal{W} = \order{(\Delta x/a)^2} \ll 1$. Hence, in the limit of small superpositions, it is impossible to verify entanglement, which is intuitively evident as both parts of the superposition acquire equal phases.

\paragraph{Medium superpositions, $\Delta x \approx a \approx d/2$.}
In this limit, the phases~\eqref{eqn:phases-similar-mass} together with the acceleration phases from the appendix yield
\begin{subequations}\label{eqn:phases-medium}\begin{align}
\Delta \phi &= \frac{G m_r m_s \tau}{\hbar a} \left(\frac{1}{2} + 0.420\,\frac{\taua}{\tau}\right)\\
\delta \phi &= -\frac{G m_r m_s \tau}{\hbar a} \left(\frac{1}{6} + 0.182\,\frac{\taua}{\tau}\right) \,.
\end{align}\end{subequations}
If we assume $\taua \ll \tau$, we find that for a detectable phase that maximizes the entanglement witness $\mathcal{W}$, according to equations~\eqref{eqn:expval}, we need
\begin{equation}\label{eqn:detectable-phase-requirement-medium}
\Delta\phi - \delta\phi \approx \frac{2}{3}\frac{G m_r m_s \tau}{\hbar a} \approx (2n+1)\, \pi \quad\quad (n \in \mathbb{N})\,.
\end{equation}

\paragraph{Large superpositions, $\Delta x \approx d$.}
This implies $a \ll \Delta x$, and based again on the phases~\eqref{eqn:phases-similar-mass} together with the acceleration phases from the appendix, including terms up to linear order in $a/\Delta x$, we have
\begin{subequations}\label{eqn:phases-large}\begin{align}
\Delta \phi &= \frac{G m_r m_s \tau}{\hbar a} \left( 1 + \frac{\pi \taua}{\tau} \sqrt{\frac{a}{2 \Delta x}}
- \left(1+ \frac{2\taua}{\tau}\right)\frac{a}{\Delta x}\right) \approx \frac{G m_r m_s \tau}{\hbar a} \\
\delta \phi &= -\frac{G m_r m_s \tau}{\hbar a} \left(1 + 1.161 \frac{\taua}{\tau}\right) \frac{a}{2\Delta x} \approx 0\,,
\end{align}\end{subequations}
where for the final approximations we consider only the leading order terms and notice that $\abs{\Delta \phi} \gg \abs{\delta \phi}$. Hence, according to equations~\eqref{eqn:expval}, we obtain a maximum of the entanglement witness $\mathcal{W}$ for
\begin{equation}\label{eqn:detectable-phase-requirement-large}
 \Delta \phi \approx \frac{G m_r m_s \tau}{\hbar a} \approx (2n+1)\, \pi \quad\quad (n \in \mathbb{N}) \,.
\end{equation}

\subsection{Closest approach and Casimir-Polder forces}
For the gravitational interaction of the two particles to dominate, it must be stronger than all other interactions between the two particles. Otherwise, the gravitationally induced phase difference will be obfuscated. For neutral particles, the most long range forces stem from the Casimir-Polder interaction. Hence, we require---as has been required in reference~\cite{boseSpinEntanglementWitness2017}---that the gravitational potential must be significantly stronger than the Casimir-Polder potential energy~\cite{casimirInfluenceRetardationLondonvan1948,emigCasimirForcesArbitrary2007}:
\begin{equation}\label{eqn:casimir}
\frac{G m_r m_s}{a} \gg \frac{23}{4\pi} \left(\frac{3}{4 \pi}\right)^2 \,  \, \frac{\hbar c \,\alpha_r \alpha_s\, m_r m_s}{\rho_r \rho_s \, a^7} \,,
\end{equation}
where $\rho_{r,s}$ are the densities of the two particles (still assuming that both particles are of almost equal size). The polarizability $\alpha = (\varepsilon^2 - 1)/(\varepsilon^2 + 2)$ can be derived from the static relative permittivity $\varepsilon$ (for non-ferromagnetic materials with relative permeability $\mu \approx 1$). In the limit $\varepsilon \to \infty$ (metals), one finds $\alpha = 1$, whereas the lowest naturally occuring permittivities for dielectrics are around 2.6 for lead(II) acetat~\cite{youngCompilationStaticDielectric1973}, limiting the polarizability to values between $0.35 \lesssim \alpha \leq 1$.

Assuming $\rho_r = \rho_s = \rho$ and $\alpha_r = \alpha_s = \alpha$, equation~\eqref{eqn:casimir} results in a limit on the distance $a$ between particles (measured center to center):
\begin{equation}
 a \gg \frac{1}{2\sqrt{\pi}} \left(\frac{3 \alpha}{\rho} \times \sqrt{\frac{23\, \hbar c}{G}} \right)^{1/3} \,.
\end{equation}
This and \eqref{eqn:detectable-phase-requirement-large} inserted into equation~\eqref{eqn:s0-limit} yields as a limit on the acceleration noise:
\begin{equation}
 S_0 \ll \frac{4}{\sqrt{\pi}\,\Delta x^2 \, (2n+1)} \left(\frac{81 \, \hbar^5 \, G^7 \rho^2}{23 \, c \, \alpha^2}\right)^{1/6} \leq \frac{4}{\sqrt{\pi}\,\Delta x^2} \left(\frac{81 \, \hbar^5 \, G^7 \rho^2}{23 \, c \, \alpha^2}\right)^{1/6} 
\end{equation}
This is an interesting result, showing that for given mass density and permittivity (which are both limited by material choices) the \emph{only} way to overcome acceleration noise is to \emph{decrease} the size of the superposition $\Delta x$. If this may sound unintuitive at first, remember that the phase uncertainty $\Delta \chi$ scales with $\Delta x$.

We did, however, learn previously that in the limit $\Delta x \ll d$ there is no entanglement. Hence, we cannot decrease $\Delta x$ much below the medium superposition regime $\Delta x \approx a \approx d/2$. Note that the factor $2/3$ in equation~\eqref{eqn:detectable-phase-requirement-medium} only further tightens the limits on $S_0$. In addition, one can easily show that with the phases~\eqref{eqn:phases-medium} for the medium superposition regime, in order to achieve $\mathcal{W} > 1$ from equations~\eqref{eqn:expval} one requires $\gamma \lesssim 0.5$.
Inserting this with \eqref{eqn:detectable-phase-requirement-medium} into \eqref{eqn:s0-limit}, using $\Delta x \approx a$, yields
\begin{equation}\label{eqn:final-noise-limit}
 S_0 \lesssim \frac{8 \, \hbar \, G}{3\pi\,a^3} \ll 
 \frac{64 \,\rho}{9\,\alpha} \sqrt{\frac{\pi \hbar G^3}{23 \,c}} \,.
\end{equation}
We found an absolute limit for the acceleration noise, depending only on the material properties (density and polarizability). Essentially, the Casimir-Polder force puts an absolute limit on the particle distance $a$; the requirement to have a detectable gravitational phase shift then requires $\Delta x \gtrsim a$ as well as $m^2 \tau$ above some limit. Hence, the phase uncertainty $\Delta \chi \sim m^2 \Delta x^2 \tau S_0$ is limited from below by the noise $S_0$ only, yielding the absolute limit for said noise.

Plugging into equation~\eqref{eqn:final-noise-limit} the values for diamond (as used in~\cite{boseSpinEntanglementWitness2017}), $\varepsilon=5.7$ and $\rho = \unit{3.5}{\gram\per\centi\meter\cubed}$, we find $\sqrt{S_0} \ll \unit{0.07}{\femto\meter\,\power{\second}{-2}/\sqrt{\hertz}}$.
This is almost two orders of magnitude below the acceleration noise of $\sqrt{S_0} \approx \unit{5.6}{\femto\meter\,\power{\second}{-2}/\sqrt{\hertz}}$ achieved by the LISA Pathfinder experiment~\cite{armanoSubfemtogFreeFall2016}.

If instead we take into account that for realistic materials $\alpha > 0.35$ and the element with the largest density is osmium with $\rho = \unit{23}{\gram\per\centi\meter\cubed}$, we get an absolute limit of
$\sqrt{S_0} \ll \unit{0.24}{\femto\meter\,\power{\second}{-2}/\sqrt{\hertz}}$.
Entanglement between two particles of equal mass can only be measured if the acceleration noise stays more than an order of magnitude below the value achieved for LISA Pathfinder.
It should be stressed again, that this represents an absolute best case scenario, which cannot be superseeded by any reasonable choice of materials, particle mass, distances, magnetic field gradients etc. The only major assumption entering our considerations which could fundamentally change this result is the similarity of the masses $m_r \approx m_s$.

\subsection{Particles of different mass in a single field gradient}

Thus far in this section, we assumed that both particles have a similar mass.
Let us now consider the opposite case, where the two particles are of very different masses, $m_r \ll m_s$. We remain, however, in the situation where a single magnetic field gradient is used to create the superpositions, in which case $\Delta x_r \gg \Delta x_s$ and, therefore,
\begin{equation}
 d > \Delta x \approx \frac{\Delta x_r}{2}
 = \frac{\mu_B \, \partial_x B \, \taua^2}{4 m_r} 
 \quad\quad\text{and}\quad\quad 
 \delta x = \Delta x + \Delta x_s \,.
\end{equation}
Then we find that $\phi^\tau_\uu \approx \phi^\tau_\ud$ and 
$\phi^\tau_\du \approx \phi^\tau_\dd$, and in the appendix it becomes clear that the same is true for the acceleration phases. With the global phase
\begin{align}
 \phi &= \frac{G m_r m_s \tau}{\hbar (d + \Delta x)} + \frac{2 G m_r m_s \taua}{\hbar d} - \Delta \phi^\text{acc}_+
 \intertext{and the phases (see appendix for the acceleration phases)}
 \Delta \phi &= \frac{G m_r m_s \tau}{\hbar (d - \Delta x)} - \frac{G m_r m_s \tau}{\hbar (d + \Delta x)} + \Delta \phi^\text{acc}_+ + \Delta \phi^\text{acc}_- \\
 \delta \phi_\uu &= \frac{G m_r m_s \tau}{\hbar (d + \delta x)} - \frac{G m_r m_s \tau}{\hbar (d + \Delta x)}  + \Delta\phi^\text{acc}_+ - \delta\phi^\text{acc}_+ = \order{\frac{\Delta x_s}{d}} \\
 \delta \phi_\dd &= \frac{G m_r m_s \tau}{\hbar (d - \delta x)} - \frac{G m_r m_s \tau}{\hbar (d - \Delta x)} - \Delta\phi^\text{acc}_- + \delta\phi^\text{acc}_-  = \order{\frac{\Delta x_s}{d}}
 \intertext{the final state reads}
 \ket{\Psi}_\tau &= \frac{\rme^{\rmi \phi}}{2} \left(\rme^{\rmi (\chi + \delta\phi_\uu)} \kuu 
 +  \kud 
 + \rme^{\rmi \Delta\phi} \kdu 
 + \rme^{\rmi(\Delta \phi + \delta \phi_\dd -\chi)} \kdd \right) \,.
\end{align}
With this, we obtain the density matrix in the basis $\{\kuu,\kud,\kdu,\kdd\}$:
\begin{equation}
 \rho 
 = \frac{1}{4} \left(\begin{array}{cccc}
1 & \rme^{-\gamma+\rmi \delta\phi_\uu} & \rme^{-\gamma - \rmi (\Delta \phi - \delta\phi_\uu)} & \rme^{-4\gamma-\rmi (\Delta\phi-\delta\phi_\uu+\delta\phi_\dd)} \\
\rme^{-\gamma-\rmi\delta\phi_\uu} & 1 & \rme^{-\rmi \Delta \phi} & \rme^{-\gamma - \rmi (\Delta \phi+\delta\phi_\dd)} \\
\rme^{-\gamma + \rmi (\Delta \phi-\delta\phi_\uu)} & \rme^{\rmi \Delta \phi} & 1 & \rme^{-\gamma-\rmi\delta\phi_\dd} \\
\rme^{-4\gamma+\rmi (\Delta\phi-\delta\phi_\uu+\delta\phi_\dd)} & \rme^{-\gamma + \rmi (\Delta \phi+\delta\phi_\dd)} & \rme^{-\gamma+\rmi\delta\phi_\dd} & 1
\end{array}\right) \,.
\end{equation}
The expectation values can then be caluclated as $\erw{\sigma_x \otimes \sigma_z}_\tau = \order{\Delta x_s/d}$ and 
\begin{equation}
\erw{\sigma_y \otimes \sigma_y}_\tau = \frac{1}{2} \left(1-\rme^{-4\gamma}\right) \cos\Delta\phi + \order{\frac{\Delta x_s}{d}} \,.
\end{equation}
This implies $\mathcal{W} \leq \frac{1}{2} +\order{\Delta x_s/d}$ (and $\mathcal{W} \approx 0$ in the absence of acceleration noise), i.\,e.\ there will be no evidence of entanglement.
Note that also $\erw{\sigma_z \otimes \sigma_x}_\tau = \order{\Delta x_s/d}$, hence the same result applies for the situation where $m_r \gg m_s$.

\subsection{Particles of different mass in equally large superpositions}\label{sec:diff-mass-equal-dx}
Obviously, the absence of entanglement in the previously discussed situation is due to the small size of $\Delta x_s$. If, alternatively, we assume two different field gradients, chosen such that $\Delta x_r \approx \Delta x_s$ despite the largely different masses $m_r \ll m_s$, we modify equation~\eqref{eqn:delta-x-dep-field-gradient} resulting in only the mass $m_s$ contributing to the phase $\chi$, which will be half as large with otherwise identical results as in the case $m_r \approx m_s$ before.
Instead of equation~\eqref{eqn:s0-limit}, we then have
\begin{equation}\label{eqn:s0-limit-diffmass}
 S_0 \lesssim \frac{32 \gamma \,\hbar^2}{m_s^2 \, \tau \, \Delta x^2} \,.
\end{equation}

Rather than from Casimi-Polder forces, the minimal approach distance is determined by the radius $R$ of the larger particle with mass $m_s$. As before, observable entanglement with the largest possible acceleration noise is achieved in the regime of medium sized superpositions $\Delta x \approx a \approx R \approx d/2$, where $\gamma \lesssim 0.5$. Hence, equation~\eqref{eqn:s0-limit-diffmass} leads to
\begin{equation}\label{eqn:s0-limit-radius}
\sqrt{S_0} \lesssim \frac{3 \,\hbar}{\pi \rho R^4 \, \sqrt{\tau}} \,,
\end{equation}
implying that smaller radius $R$ and flight time $\tau$ allow for larger acceleration noise. However, since the second particle must be smaller and $m_r m_s \tau$ sufficiently large for an observable phase, $R^4 \sqrt{\tau}$ cannot be arbitrarily small.

Firstly, the time $\tau$ must be smaller than the decoherence time from collisional decoherence~\cite{carlessoDecoherenceDueGravitational2016},
\begin{equation}\label{eqn:decoherence-time-condition}
 \tau < \frac{\sqrt{k_B T m_\text{gas}}}{16\sqrt{3}\,\zeta(3/2) P R^2} \,,
\end{equation}
where we assume a gas environment with particles of mass $m_\text{gas}$ at pressure $P$ and temperature $T$, $k_B$ being the Boltzmann constant and $\zeta$ the Riemann zeta function.
In combination with the requirement~\eqref{eqn:detectable-phase-requirement-medium} for a detectable phase we then find as a lower limit on the smaller mass $m_r$:
\begin{equation}\label{eqn:detectable-phase-and-mass-difference}
m_r \approx \frac{3}{2}\,\frac{(2n+1)\pi \, \hbar \, a}{G\,m_s\,\tau}
\geq \frac{9\, \hbar}{8\, G\,\rho\,R^2\,\tau}
> \frac{18\,\sqrt{3}\,\zeta(3/2)\,\hbar\, P}{G\,\rho\,\sqrt{k_B T m_\text{gas}}} \,.
\end{equation}
On the other hand, equation~\eqref{eqn:detectable-phase-and-mass-difference} also implies
\begin{equation}
R^4 \, \sqrt{\tau} = \frac{3}{4\pi\,\rho} \, m_s \, R \, \sqrt{\tau}
\geq \frac{9\, \sqrt{\hbar}\,m_s}{8\,\pi \sqrt{2\,G\,\rho^3\,m_r}}
\gg \frac{27\,\hbar\,\sqrt{\zeta(3/2)\, P}}{8\,\pi G\,\rho^2}\,\left(\frac{1}{3} k_B T m_\text{gas}\right)^{-1/4}
\end{equation}
where we used $m_s \gg m_r$ together with~\eqref{eqn:detectable-phase-and-mass-difference} in the last step.
Inserting this result into equation~\eqref{eqn:s0-limit-radius}, we obtain a fundamental limit for the acceleration noise:
\begin{equation}\label{s0-limit-final-diff-mass}
 \sqrt{S_0} \ll \frac{8\, G\,\rho}{9\,\sqrt{\zeta(3/2)\, P}}\,\left(\frac{1}{3} k_B T m_\text{gas}\right)^{1/4}
 = \frac{8\, G\,\rho}{9\,\sqrt{\zeta(3/2)\, n_\text{gas}}}\,\left(\frac{m_\text{gas}}{3 k_B T}\right)^{1/4}\,,
\end{equation}
where we used the ideal gas equation $P= n_\text{gas} k_B T$ with particle density $n_\text{gas}$.

As a practical example, we can insert the parameters assumed by Bose et~al.~\cite{boseSpinEntanglementWitness2017}: diamond with $\rho \approx \unit{3.5}{\gram\per\centi\meter\cubed}$, $P \approx \unit{10^{-15}}{\pascal}$, and $T \approx \unit{150}{\milli\kelvin}$, with $m_\text{gas}$ the atomic weight of nitrogen, and obtain a limit of $\unit{1.4}{\pico\meter\,\power{s}{-2}/\sqrt{\hertz}}$.

\section{Discussion}\label{sec:discussion}
We find that, although equation~\eqref{s0-limit-final-diff-mass} poses a weaker limit than~\eqref{eqn:final-noise-limit},  noise requirements are still orders of magnitude below what is usually achieved on Earth. Contrary to the limit~\eqref{eqn:final-noise-limit} for particles of similar mass, the constraint~\eqref{s0-limit-final-diff-mass} is not limited in an absolute sense by fundamental parameters and material properties. Nonetheless, even at the vacuum quality of the interstellar medium with $n_\text{gas} \sim 1/\centi\meter\cubed$ and microkelvin temperatures one would require acceleration noise below $\nano\meter\,\power{\second}{-2}/\sqrt{\hertz}$. Verifying gravitationally induced entanglement with an acceleration noise background above this value---which includes typical experiments on Earth---seems extremely challenging, if not infeasible.

Let us address some potential loopholes in our arguments. Firstly, we assumed white noise, as well as a perfectly constant, perfectly aligned field gradient $\partial_x B$ over the time and extent of the experiment. There is no obvious way in which losening these assumptions could better the situation, quite to the contrary it seems reasonable that imperfections will only result in additional noise. Stochastic fluctuations in the preparation of the experiment have been studied by Nguyen and Bernards~\cite{nguyenEntanglementDynamicsTwo2020}.

Furthermore, we took the acceleration period to be short compared with the free flight, $\taua \ll \tau$. In the opposite case, dominant contributions to both the gravitational phase $\Delta \phi$ and the noise phase $\chi$ will stem from the acceleration period rather than the free flight, and decoherence restricts $\taua$. Although calculations will become more complicated, it seems evident that our considerations remain valid, at least as far as the orders of magnitude of relevant effects are concerned.
Similarly, a deviation from the assumption that $\Delta x_r \approx \Delta x_s$ in section~\ref{sec:diff-mass-equal-dx} will not result in significantly different bounds.

Finally, rather than in a series of repeated measurements on the same particles, one could think about a set-up where one gains statistical data from an arrangement of identical experiments performed simultaneously. A time dependent external acceleration $\vec g(t)$ would only contribute to an overall phase which is the same for all measurements, although spatial fluctuations of $\vec g$ would still be required to be sufficiently small. It is beyond the scope of this article to judge the feasibility of such an idea. However, relative acceleration between the different copies would pose problems and creating copies of the experiment that are almost perfectly identical regarding particle masses, distances, and magnetic fields appears to be a tremendous challenge.

Since the distinction between unperturbed Hamiltonian and perturbation in the \schr\ equation~\eqref{eqn:schroedinger} is somewhat arbitrary (as long as the perturbation is small compared to the kinetic energy), we could of course also have considered the external acceleration as part of the unperturbed Hamiltonian $\hat{H}_0$ rather than the perturbation $\hat{\Gamma}_t$. Then we would not get the phase $\chi$ directly; however, the contribution of the external acceleration to $\hat{H}_0$ results in a (time dependent, stochastic) change of the classical trajectories, which will alter the optical path lengths and result in the same effective phase.

Our analysis focused on the specific scenario outlined by Bose et~al.~\cite{boseSpinEntanglementWitness2017}, where spin is used as an entanglement witness for gravitational interactions. The main results, however, are quite generally applicable. The precise mechanism used to create spatial superposition states is irrelevant, as long as different parts of the superposition are subject to different gravitational potentials. The decision to use spin as an entanglement witness is also merely a practical consideration: essentially the entanglement occurs purely due to the position superposition and could potentially be witnessed in any way.

Therefore, it seems safe to say that we have shown with rather general applicability that experimental attempts to witness the entanglement between two massive particles due to their gravitational interaction can only be successful in an environment with incredibly low acceleration noise. There seems no obvious route towards conducting such an experiment on Earth. Acceleration noise should play a crucial role in the evaluation of the feasibility of any possible scenario, including space missions.


\appendix
\section{Acceleration phases}
During the acceleration we collect the phases
\begin{subequations}\begin{align}
\phi^\text{acc}_\uu &= \frac{G m_r m_s}{\hbar} \left(\int_0^{\taua} \frac{\D t}{{ d +  \delta x \Sigma_\text{acc}(t)}}
+\int_{\tau+\taua}^{\tau+2\taua} \frac{\D t}{{ d +  \delta x \Sigma_\text{acc}(t)}}\right) \nnl
&= \frac{2 G m_r m_s \tau_\text{acc}}{\hbar d} f\left(\frac{\delta x}{d}\right) \\
\phi^\text{acc}_\ud &= \frac{2 G m_r m_s \tau_\text{acc}}{\hbar d} f\left(\frac{\Delta x}{d}\right) \\
\phi^\text{acc}_\du &= \frac{2 G m_r m_s \tau_\text{acc}}{\hbar d} f\left(-\frac{\Delta x}{d}\right)\\
\phi^\text{acc}_\dd &= \frac{2 G m_r m_s \tau_\text{acc}}{\hbar d} f\left(-\frac{\delta x}{d}\right)\,,
\intertext{where}
f(u) &= \int_0^{\frac{1}{2}} \frac{\D s}{{ 1 + 2 u s^2}} + \int_{\frac{1}{2}}^1 \frac{\D s}{{ 1 - 2 u s^2 + 4 u s - u}}
\quad\quad (u \in (-1,1))\,.
\end{align}\end{subequations}
For $1>u>0$ we find
\begin{equation}
f(u) = \frac{\mathrm{arctan}\left(\sqrt{\frac{u}{2}}\right)}{\sqrt{2u}} 
+ \frac{\mathrm{arctanh}\left(\sqrt{\frac{u}{2(u+1)}}\right)}{\sqrt{2u(u+1)}}   \,,
\end{equation}
and for $-1<u<0$
\begin{equation}
f(u) = \frac{\mathrm{arctan}\left(\sqrt{\frac{-u}{2(u+1)}}\right)}{\sqrt{-2u(u+1)}} 
+ \frac{\mathrm{arctanh}\left(\sqrt{\frac{-u}{2}}\right)}{\sqrt{-2u}}  \,.
\end{equation}
Expanded around $u \approx 0$, both results yield
\begin{equation}
 f(u) \approx 1 - \frac{u}{2} + \frac{23}{60} u^2 + \order{u^3} \,.
\end{equation}
The limiting case for $u \to -1$ is
\begin{equation}
f(u) \stackrel{u \to -1}{\longrightarrow} \frac{\pi}{2\sqrt{2(u+1)}} \,,
\end{equation}
and $f(u)$ takes the specific values
\begin{align}
f(-1/2) &= \sqrt{2} \,\mathrm{arctan}(1/\sqrt{2}) + \mathrm{arctanh}(1/2) \approx 1.420 \\
f(1/2) &= \mathrm{arctan}(1/2) + \sqrt{\frac{2}{3}} \,\mathrm{arctanh}(1/\sqrt{6}) \approx 0.818 \\
f(1) &= \frac{\mathrm{arctan}(1/\sqrt{2})}{\sqrt{2}} + \frac{\mathrm{arctanh(1/2)}}{2} \approx 0.710 \,.
\end{align}
We can write the phases including a global phase:
\begin{subequations}\begin{align}
\phi^\text{acc}_\uu &= \frac{2 G m_r m_s \taua}{\hbar d} - \delta\phi^\text{acc}_+ \\
\phi^\text{acc}_\ud &= \frac{2 G m_r m_s \taua}{\hbar d} - \Delta\phi^\text{acc}_+ \\
\phi^\text{acc}_\du &= \frac{2 G m_r m_s \taua}{\hbar d} + \Delta\phi^\text{acc}_- \\
\phi^\text{acc}_\dd &= \frac{2 G m_r m_s \taua}{\hbar d} + \delta\phi^\text{acc}_-
\intertext{with}
\delta\phi^\text{acc}_+ &= \frac{2 G m_r m_s \taua}{\hbar d} \left(1-f\left(\frac{\delta x}{d}\right)\right) \\
\delta\phi^\text{acc}_- &= -\frac{2 G m_r m_s \taua}{\hbar d} \left(1-f\left(-\frac{\delta x}{d}\right)\right) \\
\Delta\phi^\text{acc}_+ &= \frac{2 G m_r m_s \taua}{\hbar d} \left(1-f\left(\frac{\Delta x}{d}\right)\right) \\
\Delta\phi^\text{acc}_- &= -\frac{2 G m_r m_s \taua}{\hbar d} \left(1-f\left(-\frac{\Delta x}{d}\right)\right) \,.
\end{align}\end{subequations}
For the differences $\Delta\phi^\text{acc}_\pm$, we introduce $a = d-\Delta x$ and distinguish three cases:
\begin{enumerate}
\item For $\Delta x \ll d$ ($\Leftrightarrow a \approx d$) we have 
\begin{align}
\Delta\phi^\text{acc}_- &\approx 
\frac{G m_r m_s \taua}{\hbar d} \left(\frac{\Delta x}{d} + \frac{23 \Delta x^2}{30 d^2} + \order{(\Delta x/d)^3} \right) \nnl
&\approx \frac{G m_r m_s \taua}{\hbar a} \left( \frac{\Delta x}{a} + \frac{23 \Delta x^2}{30 a^2} \right) \\
\Delta\phi^\text{acc}_+ &\approx 
\frac{G m_r m_s \taua}{\hbar d} \left(\frac{\Delta x}{d} - \frac{23 \Delta x^2}{30 d^2} + \order{(\Delta x/d)^3} \right) \nnl
&\approx \frac{G m_r m_s \taua}{\hbar a} \left( \frac{\Delta x}{a} - \frac{23 \Delta x^2}{30 a^2} \right)  \,.
\end{align}
\item For $\Delta x \approx a \approx d/2$ we have
\begin{align}
 \Delta\phi^\text{acc}_- &\approx \frac{0.839\, G m_r m_s \taua}{\hbar d}
 \approx \frac{0.420\, G m_r m_s \taua}{\hbar a}\\
 \Delta\phi^\text{acc}_+ &\approx \frac{0.365\, G m_r m_s \taua}{\hbar d}
 \approx \frac{0.182\, G m_r m_s \taua}{\hbar a} \,.
\end{align}
\item For $\Delta x \approx d$ ($\Leftrightarrow a \ll d$) we have 
\begin{align}
 \Delta\phi^\text{acc}_- &\approx \frac{G m_r m_s \taua}{\hbar \sqrt{2d}} \left(\frac{\pi}{\sqrt{d-\Delta x}}- 2\sqrt{\frac{2}{d}}\right)\nnl
 &\approx \frac{G m_r m_s \taua}{\hbar \sqrt{2 \Delta x}} \left(\frac{\pi}{\sqrt{a}}- 2\sqrt{\frac{2}{\Delta x}}\right) \nnl
 &\approx \frac{G m_r m_s \taua}{\hbar a} \left(\frac{\pi}{\sqrt{2}}\sqrt{\frac{a}{\Delta x}}- 2 \frac{a}{\Delta x}\right) \\
 \Delta\phi^\text{acc}_+ &\approx \frac{0.580\, G m_r m_s \taua}{\hbar d}
 \approx \frac{0.580\, G m_r m_s \taua}{\hbar a} \, \frac{a}{\Delta x}\,.
\end{align}
\end{enumerate}

For the contributions to the $\kuu$ and $\kdd$ states, we distinguish the two cases of similar and very different masses:
\begin{enumerate}
 \item If we assume $\delta x \ll d$, i.\,e. the mass difference between both particles is small enough to result in a neglegible difference $\Delta x_r - \Delta x_s$ in superposition sizes with respect to the intial distance $d$ of the particles, then we find
\begin{equation}
 \delta\phi^\text{acc}_- \approx \delta\phi^\text{acc}_+ \approx \frac{G m_r m_s \taua \, \delta x}{\hbar d^2}
 + \order{(\delta x/d)^2}\,.
\end{equation}
The resulting phase from the acceleration and during flight time collected in the $\kuu$ and $\kdd$ states are then $\phi_\uu = \phi + \delta\chi$ and $\phi_\dd = \phi - \delta\chi$, where $\phi$ is the global phase~\eqref{eqn:global-phase-similar-masses} and
\begin{equation}
\delta\chi = \frac{G m_r m_s \delta x}{\hbar d^2} (\tau + \taua) \,.
\end{equation}
\item Assume now that $m_r \ll m_s$ and, therefore, $\Delta x \approx \delta x \approx \Delta x_r/2$ with $\delta x = \Delta x + \Delta x_s$. Then we have $\delta \phi^\text{acc}_\pm \approx \Delta \phi^\text{acc}_\pm + \order{\Delta x_s/d}$, and the $\kuu$ and $\kud$ as well as the $\kdu$ and $\kdd$ states each acquire the same acceleration phases.
\end{enumerate}


\begin{thebibliography}{22}%
\makeatletter
\providecommand \@ifxundefined [1]{%
 \@ifx{#1\undefined}
}%
\providecommand \@ifnum [1]{%
 \ifnum #1\expandafter \@firstoftwo
 \else \expandafter \@secondoftwo
 \fi
}%
\providecommand \@ifx [1]{%
 \ifx #1\expandafter \@firstoftwo
 \else \expandafter \@secondoftwo
 \fi
}%
\providecommand \natexlab [1]{#1}%
\providecommand \enquote  [1]{``#1''}%
\providecommand \bibnamefont  [1]{#1}%
\providecommand \bibfnamefont [1]{#1}%
\providecommand \citenamefont [1]{#1}%
\providecommand \href@noop [0]{\@secondoftwo}%
\providecommand \href [0]{\begingroup \@sanitize@url \@href}%
\providecommand \@href[1]{\@@startlink{#1}\@@href}%
\providecommand \@@href[1]{\endgroup#1\@@endlink}%
\providecommand \@sanitize@url [0]{\catcode `\\12\catcode `\$12\catcode
  `\&12\catcode `\#12\catcode `\^12\catcode `\_12\catcode `\%12\relax}%
\providecommand \@@startlink[1]{}%
\providecommand \@@endlink[0]{}%
\providecommand \url  [0]{\begingroup\@sanitize@url \@url }%
\providecommand \@url [1]{\endgroup\@href {#1}{\urlprefix }}%
\providecommand \urlprefix  [0]{URL }%
\providecommand \Eprint [0]{\href }%
\providecommand \doibase [0]{http://dx.doi.org/}%
\providecommand \selectlanguage [0]{\@gobble}%
\providecommand \bibinfo  [0]{\@secondoftwo}%
\providecommand \bibfield  [0]{\@secondoftwo}%
\providecommand \translation [1]{[#1]}%
\providecommand \BibitemOpen [0]{}%
\providecommand \bibitemStop [0]{}%
\providecommand \bibitemNoStop [0]{.\EOS\space}%
\providecommand \EOS [0]{\spacefactor3000\relax}%
\providecommand \BibitemShut  [1]{\csname bibitem#1\endcsname}%
\let\auto@bib@innerbib\@empty
\bibitem [{\citenamefont {Rosenfeld}(1963)}]{rosenfeldQuantizationFields1963}%
  \BibitemOpen
  \bibfield  {author} {\bibinfo {author} {\bibfnamefont {L.}~\bibnamefont
  {Rosenfeld}},\ }\bibfield  {title} {\enquote {\bibinfo {title} {On
  quantization of fields},}\ }\href {\doibase 10.1016/0029-5582(63)90279-7}
  {\bibfield  {journal} {\bibinfo  {journal} {Nuclear Physics}\ }\textbf
  {\bibinfo {volume} {40}},\ \bibinfo {pages} {353--356} (\bibinfo {year}
  {1963})}\BibitemShut {NoStop}%
\bibitem [{\citenamefont
  {Mattingly}(2005)}]{mattinglyQuantumGravityNecessary2005}%
  \BibitemOpen
  \bibfield  {author} {\bibinfo {author} {\bibfnamefont {James}\ \bibnamefont
  {Mattingly}},\ }\bibfield  {title} {\enquote {\bibinfo {title} {Is {{Quantum
  Gravity Necessary}}?}}\ }in\ \href@noop {} {\emph {\bibinfo {booktitle}
  {Einstein {{Studies Volume}} 11. {{The Universe}} of {{General
  Relativity}}}}},\ \bibinfo {series and number} {Einstein {{Studies}}},\
  \bibinfo {editor} {edited by\ \bibinfo {editor} {\bibfnamefont {A.~J.}\
  \bibnamefont {Kox}}\ and\ \bibinfo {editor} {\bibfnamefont {Jean}\
  \bibnamefont {Eisenstaedt}}}\ (\bibinfo  {publisher} {{Birkh\"auser}},\
  \bibinfo {address} {{Boston}},\ \bibinfo {year} {2005})\ pp.\ \bibinfo
  {pages} {327--338}\BibitemShut {NoStop}%
\bibitem [{\citenamefont {Eppley}\ and\ \citenamefont
  {Hannah}(1977)}]{eppleyNecessityQuantizingGravitational1977}%
  \BibitemOpen
  \bibfield  {author} {\bibinfo {author} {\bibfnamefont {Kenneth}\ \bibnamefont
  {Eppley}}\ and\ \bibinfo {author} {\bibfnamefont {Eric}\ \bibnamefont
  {Hannah}},\ }\bibfield  {title} {\enquote {\bibinfo {title} {The necessity of
  quantizing the gravitational field},}\ }\href {\doibase 10.1007/BF00715241}
  {\bibfield  {journal} {\bibinfo  {journal} {Foundations of Physics}\ }\textbf
  {\bibinfo {volume} {7}},\ \bibinfo {pages} {51--68} (\bibinfo {year}
  {1977})}\BibitemShut {NoStop}%
\bibitem [{\citenamefont {Page}\ and\ \citenamefont
  {Geilker}(1981)}]{pageIndirectEvidenceQuantum1981}%
  \BibitemOpen
  \bibfield  {author} {\bibinfo {author} {\bibfnamefont {Don~N.}\ \bibnamefont
  {Page}}\ and\ \bibinfo {author} {\bibfnamefont {C.~D.}\ \bibnamefont
  {Geilker}},\ }\bibfield  {title} {\enquote {\bibinfo {title} {Indirect
  {{Evidence}} for {{Quantum Gravity}}},}\ }\href {\doibase
  10.1103/PhysRevLett.47.979} {\bibfield  {journal} {\bibinfo  {journal}
  {Physical Review Letters}\ }\textbf {\bibinfo {volume} {47}},\ \bibinfo
  {pages} {979--982} (\bibinfo {year} {1981})}\BibitemShut {NoStop}%
\bibitem [{\citenamefont {Albers}\ \emph {et~al.}(2008)\citenamefont {Albers},
  \citenamefont {Kiefer},\ and\ \citenamefont
  {Reginatto}}]{albersMeasurementAnalysisQuantum2008}%
  \BibitemOpen
  \bibfield  {author} {\bibinfo {author} {\bibfnamefont {Mark}\ \bibnamefont
  {Albers}}, \bibinfo {author} {\bibfnamefont {Claus}\ \bibnamefont {Kiefer}},
  \ and\ \bibinfo {author} {\bibfnamefont {Marcel}\ \bibnamefont {Reginatto}},\
  }\bibfield  {title} {\enquote {\bibinfo {title} {Measurement analysis and
  quantum gravity},}\ }\href {\doibase 10.1103/PhysRevD.78.064051} {\bibfield
  {journal} {\bibinfo  {journal} {Physical Review D}\ }\textbf {\bibinfo
  {volume} {78}},\ \bibinfo {pages} {064051} (\bibinfo {year}
  {2008})}\BibitemShut {NoStop}%
\bibitem [{\citenamefont {Carlip}(2008)}]{carlipQuantumGravityNecessary2008}%
  \BibitemOpen
  \bibfield  {author} {\bibinfo {author} {\bibfnamefont {S.}~\bibnamefont
  {Carlip}},\ }\bibfield  {title} {\enquote {\bibinfo {title} {Is quantum
  gravity necessary?}}\ }\href {\doibase 10.1088/0264-9381/25/15/154010}
  {\bibfield  {journal} {\bibinfo  {journal} {Classical and Quantum Gravity}\
  }\textbf {\bibinfo {volume} {25}},\ \bibinfo {pages} {154010} (\bibinfo
  {year} {2008})}\BibitemShut {NoStop}%
\bibitem [{\citenamefont {Yang}\ \emph {et~al.}(2013)\citenamefont {Yang},
  \citenamefont {Miao}, \citenamefont {Lee}, \citenamefont {Helou},\ and\
  \citenamefont {Chen}}]{yangMacroscopicQuantumMechanics2013}%
  \BibitemOpen
  \bibfield  {author} {\bibinfo {author} {\bibfnamefont {Huan}\ \bibnamefont
  {Yang}}, \bibinfo {author} {\bibfnamefont {Haixing}\ \bibnamefont {Miao}},
  \bibinfo {author} {\bibfnamefont {Da-Shin}\ \bibnamefont {Lee}}, \bibinfo
  {author} {\bibfnamefont {Bassam}\ \bibnamefont {Helou}}, \ and\ \bibinfo
  {author} {\bibfnamefont {Yanbei}\ \bibnamefont {Chen}},\ }\bibfield  {title}
  {\enquote {\bibinfo {title} {Macroscopic {{Quantum Mechanics}} in a
  {{Classical Spacetime}}},}\ }\href {\doibase 10.1103/PhysRevLett.110.170401}
  {\bibfield  {journal} {\bibinfo  {journal} {Physical Review Letters}\
  }\textbf {\bibinfo {volume} {110}},\ \bibinfo {pages} {170401} (\bibinfo
  {year} {2013})}\BibitemShut {NoStop}%
\bibitem [{\citenamefont {Gro{\ss}ardt}\ \emph {et~al.}(2016)\citenamefont
  {Gro{\ss}ardt}, \citenamefont {Bateman}, \citenamefont {Ulbricht},\ and\
  \citenamefont {Bassi}}]{grossardtOptomechanicalTestSchrodingerNewton2016}%
  \BibitemOpen
  \bibfield  {author} {\bibinfo {author} {\bibfnamefont {Andr{\'e}}\
  \bibnamefont {Gro{\ss}ardt}}, \bibinfo {author} {\bibfnamefont {James}\
  \bibnamefont {Bateman}}, \bibinfo {author} {\bibfnamefont {Hendrik}\
  \bibnamefont {Ulbricht}}, \ and\ \bibinfo {author} {\bibfnamefont {Angelo}\
  \bibnamefont {Bassi}},\ }\bibfield  {title} {\enquote {\bibinfo {title}
  {Optomechanical test of the {{Schr\"odinger}}-{{Newton}} equation},}\ }\href
  {\doibase 10.1103/PhysRevD.93.096003} {\bibfield  {journal} {\bibinfo
  {journal} {Physical Review D}\ }\textbf {\bibinfo {volume} {93}},\ \bibinfo
  {pages} {096003} (\bibinfo {year} {2016})}\BibitemShut {NoStop}%
\bibitem [{\citenamefont
  {M{\o}ller}(1962)}]{mollerTheoriesRelativistesGravitation1962}%
  \BibitemOpen
  \bibfield  {author} {\bibinfo {author} {\bibfnamefont {C.}~\bibnamefont
  {M{\o}ller}},\ }\bibfield  {title} {\enquote {\bibinfo {title} {{Les
  Th\'eories Relativistes de la Gravitation}},}\ }in\ \href@noop {} {\emph
  {\bibinfo {booktitle} {{Colloques Internationaux CNRS}}}},\ Vol.~\bibinfo
  {volume} {91},\ \bibinfo {editor} {edited by\ \bibinfo {editor}
  {\bibfnamefont {A.}~\bibnamefont {Lichnerowicz}}\ and\ \bibinfo {editor}
  {\bibfnamefont {M.-A.}\ \bibnamefont {Tonnelat}}}\ (\bibinfo  {publisher}
  {{CNRS, Paris}},\ \bibinfo {year} {1962})\BibitemShut {NoStop}%
\bibitem [{\citenamefont {Bahrami}\ \emph {et~al.}(2014)\citenamefont
  {Bahrami}, \citenamefont {Gro{\ss}ardt}, \citenamefont {Donadi},\ and\
  \citenamefont {Bassi}}]{bahramiSchrodingerNewtonEquationIts2014}%
  \BibitemOpen
  \bibfield  {author} {\bibinfo {author} {\bibfnamefont {Mohammad}\
  \bibnamefont {Bahrami}}, \bibinfo {author} {\bibfnamefont {Andr{\'e}}\
  \bibnamefont {Gro{\ss}ardt}}, \bibinfo {author} {\bibfnamefont {Sandro}\
  \bibnamefont {Donadi}}, \ and\ \bibinfo {author} {\bibfnamefont {Angelo}\
  \bibnamefont {Bassi}},\ }\bibfield  {title} {\enquote {\bibinfo {title} {The
  {{Schr\"odinger}}-{{Newton}} equation and its foundations},}\ }\href
  {\doibase 10.1088/1367-2630/16/11/115007} {\bibfield  {journal} {\bibinfo
  {journal} {New Journal of Physics}\ }\textbf {\bibinfo {volume} {16}},\
  \bibinfo {pages} {115007} (\bibinfo {year} {2014})}\BibitemShut {NoStop}%
\bibitem [{\citenamefont {Schm{\"o}le}\ \emph {et~al.}(2016)\citenamefont
  {Schm{\"o}le}, \citenamefont {Dragosits}, \citenamefont {Hepach},\ and\
  \citenamefont
  {Aspelmeyer}}]{schmoleMicromechanicalProofofprincipleExperiment2016}%
  \BibitemOpen
  \bibfield  {author} {\bibinfo {author} {\bibfnamefont {Jonas}\ \bibnamefont
  {Schm{\"o}le}}, \bibinfo {author} {\bibfnamefont {Mathias}\ \bibnamefont
  {Dragosits}}, \bibinfo {author} {\bibfnamefont {Hans}\ \bibnamefont
  {Hepach}}, \ and\ \bibinfo {author} {\bibfnamefont {Markus}\ \bibnamefont
  {Aspelmeyer}},\ }\bibfield  {title} {\enquote {\bibinfo {title} {A
  micromechanical proof-of-principle experiment for measuring the gravitational
  force of milligram masses},}\ }\href {\doibase
  10.1088/0264-9381/33/12/125031} {\bibfield  {journal} {\bibinfo  {journal}
  {Classical and Quantum Gravity}\ }\textbf {\bibinfo {volume} {33}},\ \bibinfo
  {pages} {125031} (\bibinfo {year} {2016})}\BibitemShut {NoStop}%
\bibitem [{\citenamefont {Bose}\ \emph {et~al.}(2017)\citenamefont {Bose},
  \citenamefont {Mazumdar}, \citenamefont {Morley}, \citenamefont {Ulbricht},
  \citenamefont {Toro{\v s}}, \citenamefont {Paternostro}, \citenamefont
  {Geraci}, \citenamefont {Barker}, \citenamefont {Kim},\ and\ \citenamefont
  {Milburn}}]{boseSpinEntanglementWitness2017}%
  \BibitemOpen
  \bibfield  {author} {\bibinfo {author} {\bibfnamefont {Sougato}\ \bibnamefont
  {Bose}}, \bibinfo {author} {\bibfnamefont {Amupam}\ \bibnamefont {Mazumdar}},
  \bibinfo {author} {\bibfnamefont {Gavin~W.}\ \bibnamefont {Morley}}, \bibinfo
  {author} {\bibfnamefont {Hendrik}\ \bibnamefont {Ulbricht}}, \bibinfo
  {author} {\bibfnamefont {Marko}\ \bibnamefont {Toro{\v s}}}, \bibinfo
  {author} {\bibfnamefont {Mauro}\ \bibnamefont {Paternostro}}, \bibinfo
  {author} {\bibfnamefont {Andrew~A.}\ \bibnamefont {Geraci}}, \bibinfo
  {author} {\bibfnamefont {Peter~F.}\ \bibnamefont {Barker}}, \bibinfo {author}
  {\bibfnamefont {M.~S.}\ \bibnamefont {Kim}}, \ and\ \bibinfo {author}
  {\bibfnamefont {Gerard}\ \bibnamefont {Milburn}},\ }\bibfield  {title}
  {\enquote {\bibinfo {title} {Spin {{Entanglement Witness}} for {{Quantum
  Gravity}}},}\ }\href {\doibase 10.1103/PhysRevLett.119.240401} {\bibfield
  {journal} {\bibinfo  {journal} {Physical Review Letters}\ }\textbf {\bibinfo
  {volume} {119}},\ \bibinfo {pages} {240401} (\bibinfo {year}
  {2017})}\BibitemShut {NoStop}%
\bibitem [{\citenamefont {Colella}\ \emph {et~al.}(1975)\citenamefont
  {Colella}, \citenamefont {Overhauser},\ and\ \citenamefont
  {Werner}}]{colellaObservationGravitationallyInduced1975}%
  \BibitemOpen
  \bibfield  {author} {\bibinfo {author} {\bibfnamefont {R.}~\bibnamefont
  {Colella}}, \bibinfo {author} {\bibfnamefont {A.~W.}\ \bibnamefont
  {Overhauser}}, \ and\ \bibinfo {author} {\bibfnamefont {S.~A.}\ \bibnamefont
  {Werner}},\ }\bibfield  {title} {\enquote {\bibinfo {title} {Observation of
  {{Gravitationally Induced Quantum Interference}}},}\ }\href {\doibase
  10.1103/PhysRevLett.34.1472} {\bibfield  {journal} {\bibinfo  {journal}
  {Physical Review Letters}\ }\textbf {\bibinfo {volume} {34}},\ \bibinfo
  {pages} {1472--1474} (\bibinfo {year} {1975})}\BibitemShut {NoStop}%
\bibitem [{\citenamefont
  {Greenberger}(1983)}]{greenbergerNeutronInterferometerDevice1983}%
  \BibitemOpen
  \bibfield  {author} {\bibinfo {author} {\bibfnamefont {Daniel~M.}\
  \bibnamefont {Greenberger}},\ }\bibfield  {title} {\enquote {\bibinfo {title}
  {The neutron interferometer as a device for illustrating the strange behavior
  of quantum systems},}\ }\href {\doibase 10.1103/RevModPhys.55.875} {\bibfield
   {journal} {\bibinfo  {journal} {Reviews of Modern Physics}\ }\textbf
  {\bibinfo {volume} {55}},\ \bibinfo {pages} {875--905} (\bibinfo {year}
  {1983})}\BibitemShut {NoStop}%
\bibitem [{\citenamefont {Selig}\ \emph {et~al.}(2010)\citenamefont {Selig},
  \citenamefont {Dittus},\ and\ \citenamefont
  {L{\"a}mmerzahl}}]{seligDropTowerMicrogravity2010}%
  \BibitemOpen
  \bibfield  {author} {\bibinfo {author} {\bibfnamefont {Hanns}\ \bibnamefont
  {Selig}}, \bibinfo {author} {\bibfnamefont {Hansj{\"o}rg}\ \bibnamefont
  {Dittus}}, \ and\ \bibinfo {author} {\bibfnamefont {Claus}\ \bibnamefont
  {L{\"a}mmerzahl}},\ }\bibfield  {title} {\enquote {\bibinfo {title} {Drop
  {{Tower Microgravity Improvement Towards}} the {{Nano}}-g {{Level}} for the
  {{MICROSCOPE Payload Tests}}},}\ }\href {\doibase 10.1007/s12217-010-9210-0}
  {\bibfield  {journal} {\bibinfo  {journal} {Microgravity Science and
  Technology}\ }\textbf {\bibinfo {volume} {22}},\ \bibinfo {pages} {539--549}
  (\bibinfo {year} {2010})}\BibitemShut {NoStop}%
\bibitem [{\citenamefont {Armano}\ \emph {et~al.}(2016)\citenamefont {Armano},
  \citenamefont {Audley}, \citenamefont {Auger}, \citenamefont {Baird},
  \citenamefont {Bassan}, \citenamefont {Binetruy}, \citenamefont {Born},
  \citenamefont {Bortoluzzi}, \citenamefont {Brandt}, \citenamefont {Caleno},
  \citenamefont {Carbone}, \citenamefont {Cavalleri}, \citenamefont {Cesarini},
  \citenamefont {Ciani}, \citenamefont {Congedo}, \citenamefont {Cruise},
  \citenamefont {Danzmann}, \citenamefont {{de Deus Silva}}, \citenamefont
  {De~Rosa}, \citenamefont {{Diaz-Aguil{\'o}}}, \citenamefont {Di~Fiore},
  \citenamefont {Diepholz}, \citenamefont {Dixon}, \citenamefont {Dolesi},
  \citenamefont {Dunbar}, \citenamefont {Ferraioli}, \citenamefont {Ferroni},
  \citenamefont {Fichter}, \citenamefont {Fitzsimons}, \citenamefont
  {Flatscher}, \citenamefont {Freschi}, \citenamefont {Garc{\'i}a~Mar{\'i}n},
  \citenamefont {Garc{\'i}a~Marirrodriga}, \citenamefont {Gerndt},
  \citenamefont {Gesa}, \citenamefont {Gibert}, \citenamefont {Giardini},
  \citenamefont {Giusteri}, \citenamefont {Guzm{\'a}n}, \citenamefont {Grado},
  \citenamefont {Grimani}, \citenamefont {Grynagier}, \citenamefont
  {Grzymisch}, \citenamefont {Harrison}, \citenamefont {Heinzel}, \citenamefont
  {Hewitson}, \citenamefont {Hollington}, \citenamefont {Hoyland},
  \citenamefont {Hueller}, \citenamefont {Inchausp{\'e}}, \citenamefont
  {Jennrich}, \citenamefont {Jetzer}, \citenamefont {Johann}, \citenamefont
  {Johlander}, \citenamefont {Karnesis}, \citenamefont {Kaune}, \citenamefont
  {Korsakova}, \citenamefont {Killow}, \citenamefont {Lobo}, \citenamefont
  {Lloro}, \citenamefont {Liu}, \citenamefont {{L{\'o}pez-Zaragoza}},
  \citenamefont {Maarschalkerweerd}, \citenamefont {Mance}, \citenamefont
  {Mart{\'i}n}, \citenamefont {{Martin-Polo}}, \citenamefont {Martino},
  \citenamefont {{Martin-Porqueras}}, \citenamefont {Madden}, \citenamefont
  {Mateos}, \citenamefont {McNamara}, \citenamefont {Mendes}, \citenamefont
  {Mendes}, \citenamefont {Monsky}, \citenamefont {Nicolodi}, \citenamefont
  {Nofrarias}, \citenamefont {Paczkowski}, \citenamefont {{Perreur-Lloyd}},
  \citenamefont {Petiteau}, \citenamefont {Pivato}, \citenamefont {Plagnol},
  \citenamefont {Prat}, \citenamefont {Ragnit}, \citenamefont {Ra{\"i}s},
  \citenamefont {{Ramos-Castro}}, \citenamefont {Reiche}, \citenamefont
  {Robertson}, \citenamefont {Rozemeijer}, \citenamefont {Rivas}, \citenamefont
  {Russano}, \citenamefont {Sanju{\'a}n}, \citenamefont {Sarra}, \citenamefont
  {Schleicher}, \citenamefont {Shaul}, \citenamefont {Slutsky}, \citenamefont
  {Sopuerta}, \citenamefont {Stanga}, \citenamefont {Steier}, \citenamefont
  {Sumner}, \citenamefont {Texier}, \citenamefont {Thorpe}, \citenamefont
  {Trenkel}, \citenamefont {Tr{\"o}bs}, \citenamefont {Tu}, \citenamefont
  {Vetrugno}, \citenamefont {Vitale}, \citenamefont {Wand}, \citenamefont
  {Wanner}, \citenamefont {Ward}, \citenamefont {Warren}, \citenamefont {Wass},
  \citenamefont {Wealthy}, \citenamefont {Weber}, \citenamefont {Wissel},
  \citenamefont {Wittchen}, \citenamefont {Zambotti}, \citenamefont {Zanoni},
  \citenamefont {Ziegler},\ and\ \citenamefont
  {Zweifel}}]{armanoSubfemtogFreeFall2016}%
  \BibitemOpen
  \bibfield  {author} {\bibinfo {author} {\bibfnamefont {M.}~\bibnamefont
  {Armano}}, \bibinfo {author} {\bibfnamefont {H.}~\bibnamefont {Audley}},
  \bibinfo {author} {\bibfnamefont {G.}~\bibnamefont {Auger}}, \bibinfo
  {author} {\bibfnamefont {J.~T.}\ \bibnamefont {Baird}}, \bibinfo {author}
  {\bibfnamefont {M.}~\bibnamefont {Bassan}}, \bibinfo {author} {\bibfnamefont
  {P.}~\bibnamefont {Binetruy}}, \bibinfo {author} {\bibfnamefont
  {M.}~\bibnamefont {Born}}, \bibinfo {author} {\bibfnamefont {D.}~\bibnamefont
  {Bortoluzzi}}, \bibinfo {author} {\bibfnamefont {N.}~\bibnamefont {Brandt}},
  \bibinfo {author} {\bibfnamefont {M.}~\bibnamefont {Caleno}}, \bibinfo
  {author} {\bibfnamefont {L.}~\bibnamefont {Carbone}}, \bibinfo {author}
  {\bibfnamefont {A.}~\bibnamefont {Cavalleri}}, \bibinfo {author}
  {\bibfnamefont {A.}~\bibnamefont {Cesarini}}, \bibinfo {author}
  {\bibfnamefont {G.}~\bibnamefont {Ciani}}, \bibinfo {author} {\bibfnamefont
  {G.}~\bibnamefont {Congedo}}, \bibinfo {author} {\bibfnamefont {A.~M.}\
  \bibnamefont {Cruise}}, \bibinfo {author} {\bibfnamefont {K.}~\bibnamefont
  {Danzmann}}, \bibinfo {author} {\bibfnamefont {M.}~\bibnamefont {{de Deus
  Silva}}}, \bibinfo {author} {\bibfnamefont {R.}~\bibnamefont {De~Rosa}},
  \bibinfo {author} {\bibfnamefont {M.}~\bibnamefont {{Diaz-Aguil{\'o}}}},
  \bibinfo {author} {\bibfnamefont {L.}~\bibnamefont {Di~Fiore}}, \bibinfo
  {author} {\bibfnamefont {I.}~\bibnamefont {Diepholz}}, \bibinfo {author}
  {\bibfnamefont {G.}~\bibnamefont {Dixon}}, \bibinfo {author} {\bibfnamefont
  {R.}~\bibnamefont {Dolesi}}, \bibinfo {author} {\bibfnamefont
  {N.}~\bibnamefont {Dunbar}}, \bibinfo {author} {\bibfnamefont
  {L.}~\bibnamefont {Ferraioli}}, \bibinfo {author} {\bibfnamefont
  {V.}~\bibnamefont {Ferroni}}, \bibinfo {author} {\bibfnamefont
  {W.}~\bibnamefont {Fichter}}, \bibinfo {author} {\bibfnamefont {E.~D.}\
  \bibnamefont {Fitzsimons}}, \bibinfo {author} {\bibfnamefont
  {R.}~\bibnamefont {Flatscher}}, \bibinfo {author} {\bibfnamefont
  {M.}~\bibnamefont {Freschi}}, \bibinfo {author} {\bibfnamefont {A.~F.}\
  \bibnamefont {Garc{\'i}a~Mar{\'i}n}}, \bibinfo {author} {\bibfnamefont
  {C.}~\bibnamefont {Garc{\'i}a~Marirrodriga}}, \bibinfo {author}
  {\bibfnamefont {R.}~\bibnamefont {Gerndt}}, \bibinfo {author} {\bibfnamefont
  {L.}~\bibnamefont {Gesa}}, \bibinfo {author} {\bibfnamefont {F.}~\bibnamefont
  {Gibert}}, \bibinfo {author} {\bibfnamefont {D.}~\bibnamefont {Giardini}},
  \bibinfo {author} {\bibfnamefont {R.}~\bibnamefont {Giusteri}}, \bibinfo
  {author} {\bibfnamefont {F.}~\bibnamefont {Guzm{\'a}n}}, \bibinfo {author}
  {\bibfnamefont {A.}~\bibnamefont {Grado}}, \bibinfo {author} {\bibfnamefont
  {C.}~\bibnamefont {Grimani}}, \bibinfo {author} {\bibfnamefont
  {A.}~\bibnamefont {Grynagier}}, \bibinfo {author} {\bibfnamefont
  {J.}~\bibnamefont {Grzymisch}}, \bibinfo {author} {\bibfnamefont
  {I.}~\bibnamefont {Harrison}}, \bibinfo {author} {\bibfnamefont
  {G.}~\bibnamefont {Heinzel}}, \bibinfo {author} {\bibfnamefont
  {M.}~\bibnamefont {Hewitson}}, \bibinfo {author} {\bibfnamefont
  {D.}~\bibnamefont {Hollington}}, \bibinfo {author} {\bibfnamefont
  {D.}~\bibnamefont {Hoyland}}, \bibinfo {author} {\bibfnamefont
  {M.}~\bibnamefont {Hueller}}, \bibinfo {author} {\bibfnamefont
  {H.}~\bibnamefont {Inchausp{\'e}}}, \bibinfo {author} {\bibfnamefont
  {O.}~\bibnamefont {Jennrich}}, \bibinfo {author} {\bibfnamefont
  {P.}~\bibnamefont {Jetzer}}, \bibinfo {author} {\bibfnamefont
  {U.}~\bibnamefont {Johann}}, \bibinfo {author} {\bibfnamefont
  {B.}~\bibnamefont {Johlander}}, \bibinfo {author} {\bibfnamefont
  {N.}~\bibnamefont {Karnesis}}, \bibinfo {author} {\bibfnamefont
  {B.}~\bibnamefont {Kaune}}, \bibinfo {author} {\bibfnamefont
  {N.}~\bibnamefont {Korsakova}}, \bibinfo {author} {\bibfnamefont {C.~J.}\
  \bibnamefont {Killow}}, \bibinfo {author} {\bibfnamefont {J.~A.}\
  \bibnamefont {Lobo}}, \bibinfo {author} {\bibfnamefont {I.}~\bibnamefont
  {Lloro}}, \bibinfo {author} {\bibfnamefont {L.}~\bibnamefont {Liu}}, \bibinfo
  {author} {\bibfnamefont {J.~P.}\ \bibnamefont {{L{\'o}pez-Zaragoza}}},
  \bibinfo {author} {\bibfnamefont {R.}~\bibnamefont {Maarschalkerweerd}},
  \bibinfo {author} {\bibfnamefont {D.}~\bibnamefont {Mance}}, \bibinfo
  {author} {\bibfnamefont {V.}~\bibnamefont {Mart{\'i}n}}, \bibinfo {author}
  {\bibfnamefont {L.}~\bibnamefont {{Martin-Polo}}}, \bibinfo {author}
  {\bibfnamefont {J.}~\bibnamefont {Martino}}, \bibinfo {author} {\bibfnamefont
  {F.}~\bibnamefont {{Martin-Porqueras}}}, \bibinfo {author} {\bibfnamefont
  {S.}~\bibnamefont {Madden}}, \bibinfo {author} {\bibfnamefont
  {I.}~\bibnamefont {Mateos}}, \bibinfo {author} {\bibfnamefont {P.~W.}\
  \bibnamefont {McNamara}}, \bibinfo {author} {\bibfnamefont {J.}~\bibnamefont
  {Mendes}}, \bibinfo {author} {\bibfnamefont {L.}~\bibnamefont {Mendes}},
  \bibinfo {author} {\bibfnamefont {A.}~\bibnamefont {Monsky}}, \bibinfo
  {author} {\bibfnamefont {D.}~\bibnamefont {Nicolodi}}, \bibinfo {author}
  {\bibfnamefont {M.}~\bibnamefont {Nofrarias}}, \bibinfo {author}
  {\bibfnamefont {S.}~\bibnamefont {Paczkowski}}, \bibinfo {author}
  {\bibfnamefont {M.}~\bibnamefont {{Perreur-Lloyd}}}, \bibinfo {author}
  {\bibfnamefont {A.}~\bibnamefont {Petiteau}}, \bibinfo {author}
  {\bibfnamefont {P.}~\bibnamefont {Pivato}}, \bibinfo {author} {\bibfnamefont
  {E.}~\bibnamefont {Plagnol}}, \bibinfo {author} {\bibfnamefont
  {P.}~\bibnamefont {Prat}}, \bibinfo {author} {\bibfnamefont {U.}~\bibnamefont
  {Ragnit}}, \bibinfo {author} {\bibfnamefont {B.}~\bibnamefont {Ra{\"i}s}},
  \bibinfo {author} {\bibfnamefont {J.}~\bibnamefont {{Ramos-Castro}}},
  \bibinfo {author} {\bibfnamefont {J.}~\bibnamefont {Reiche}}, \bibinfo
  {author} {\bibfnamefont {D.~I.}\ \bibnamefont {Robertson}}, \bibinfo {author}
  {\bibfnamefont {H.}~\bibnamefont {Rozemeijer}}, \bibinfo {author}
  {\bibfnamefont {F.}~\bibnamefont {Rivas}}, \bibinfo {author} {\bibfnamefont
  {G.}~\bibnamefont {Russano}}, \bibinfo {author} {\bibfnamefont
  {J.}~\bibnamefont {Sanju{\'a}n}}, \bibinfo {author} {\bibfnamefont
  {P.}~\bibnamefont {Sarra}}, \bibinfo {author} {\bibfnamefont
  {A.}~\bibnamefont {Schleicher}}, \bibinfo {author} {\bibfnamefont
  {D.}~\bibnamefont {Shaul}}, \bibinfo {author} {\bibfnamefont
  {J.}~\bibnamefont {Slutsky}}, \bibinfo {author} {\bibfnamefont {C.~F.}\
  \bibnamefont {Sopuerta}}, \bibinfo {author} {\bibfnamefont {R.}~\bibnamefont
  {Stanga}}, \bibinfo {author} {\bibfnamefont {F.}~\bibnamefont {Steier}},
  \bibinfo {author} {\bibfnamefont {T.}~\bibnamefont {Sumner}}, \bibinfo
  {author} {\bibfnamefont {D.}~\bibnamefont {Texier}}, \bibinfo {author}
  {\bibfnamefont {J.~I.}\ \bibnamefont {Thorpe}}, \bibinfo {author}
  {\bibfnamefont {C.}~\bibnamefont {Trenkel}}, \bibinfo {author} {\bibfnamefont
  {M.}~\bibnamefont {Tr{\"o}bs}}, \bibinfo {author} {\bibfnamefont {H.~B.}\
  \bibnamefont {Tu}}, \bibinfo {author} {\bibfnamefont {D.}~\bibnamefont
  {Vetrugno}}, \bibinfo {author} {\bibfnamefont {S.}~\bibnamefont {Vitale}},
  \bibinfo {author} {\bibfnamefont {V.}~\bibnamefont {Wand}}, \bibinfo {author}
  {\bibfnamefont {G.}~\bibnamefont {Wanner}}, \bibinfo {author} {\bibfnamefont
  {H.}~\bibnamefont {Ward}}, \bibinfo {author} {\bibfnamefont {C.}~\bibnamefont
  {Warren}}, \bibinfo {author} {\bibfnamefont {P.~J.}\ \bibnamefont {Wass}},
  \bibinfo {author} {\bibfnamefont {D.}~\bibnamefont {Wealthy}}, \bibinfo
  {author} {\bibfnamefont {W.~J.}\ \bibnamefont {Weber}}, \bibinfo {author}
  {\bibfnamefont {L.}~\bibnamefont {Wissel}}, \bibinfo {author} {\bibfnamefont
  {A.}~\bibnamefont {Wittchen}}, \bibinfo {author} {\bibfnamefont
  {A.}~\bibnamefont {Zambotti}}, \bibinfo {author} {\bibfnamefont
  {C.}~\bibnamefont {Zanoni}}, \bibinfo {author} {\bibfnamefont
  {T.}~\bibnamefont {Ziegler}}, \ and\ \bibinfo {author} {\bibfnamefont
  {{and}~P.}\ \bibnamefont {Zweifel}},\ }\bibfield  {title} {\enquote {\bibinfo
  {title} {Sub-femto-g free fall for space-based gravitational wave
  observatories: {{LISA}} pathfinder results},}\ }\href {\doibase
  10.1103/PhysRevLett.116.231101} {\bibfield  {journal} {\bibinfo  {journal}
  {Physical Review Letters}\ }\textbf {\bibinfo {volume} {116}},\ \bibinfo
  {pages} {231101} (\bibinfo {year} {2016})}\BibitemShut {NoStop}%
\bibitem [{\citenamefont {Lamine}\ \emph {et~al.}(2002)\citenamefont {Lamine},
  \citenamefont {Jaekel},\ and\ \citenamefont
  {Reynaud}}]{lamineGravitationalDecoherenceAtomic2002}%
  \BibitemOpen
  \bibfield  {author} {\bibinfo {author} {\bibfnamefont {B.}~\bibnamefont
  {Lamine}}, \bibinfo {author} {\bibfnamefont {M.-T.}\ \bibnamefont {Jaekel}},
  \ and\ \bibinfo {author} {\bibfnamefont {S.}~\bibnamefont {Reynaud}},\
  }\bibfield  {title} {\enquote {\bibinfo {title} {Gravitational decoherence of
  atomic interferometers},}\ }\href {\doibase 10.1140/epjd/e2002-00126-y}
  {\bibfield  {journal} {\bibinfo  {journal} {The European Physical Journal D -
  Atomic, Molecular, Optical and Plasma Physics}\ }\textbf {\bibinfo {volume}
  {20}},\ \bibinfo {pages} {165--176} (\bibinfo {year} {2002})}\BibitemShut
  {NoStop}%
\bibitem [{\citenamefont {Casimir}\ and\ \citenamefont
  {Polder}(1948)}]{casimirInfluenceRetardationLondonvan1948}%
  \BibitemOpen
  \bibfield  {author} {\bibinfo {author} {\bibfnamefont {H.~B.~G.}\
  \bibnamefont {Casimir}}\ and\ \bibinfo {author} {\bibfnamefont
  {D.}~\bibnamefont {Polder}},\ }\bibfield  {title} {\enquote {\bibinfo {title}
  {The {{Influence}} of {{Retardation}} on the {{London}}-van der {{Waals
  Forces}}},}\ }\href {\doibase 10.1103/PhysRev.73.360} {\bibfield  {journal}
  {\bibinfo  {journal} {Physical Review}\ }\textbf {\bibinfo {volume} {73}},\
  \bibinfo {pages} {360--372} (\bibinfo {year} {1948})}\BibitemShut {NoStop}%
\bibitem [{\citenamefont {Emig}\ \emph {et~al.}(2007)\citenamefont {Emig},
  \citenamefont {Graham}, \citenamefont {Jaffe},\ and\ \citenamefont
  {Kardar}}]{emigCasimirForcesArbitrary2007}%
  \BibitemOpen
  \bibfield  {author} {\bibinfo {author} {\bibfnamefont {T.}~\bibnamefont
  {Emig}}, \bibinfo {author} {\bibfnamefont {N.}~\bibnamefont {Graham}},
  \bibinfo {author} {\bibfnamefont {R.~L.}\ \bibnamefont {Jaffe}}, \ and\
  \bibinfo {author} {\bibfnamefont {M.}~\bibnamefont {Kardar}},\ }\bibfield
  {title} {\enquote {\bibinfo {title} {Casimir {{Forces}} between {{Arbitrary
  Compact Objects}}},}\ }\href {\doibase 10.1103/PhysRevLett.99.170403}
  {\bibfield  {journal} {\bibinfo  {journal} {Physical Review Letters}\
  }\textbf {\bibinfo {volume} {99}},\ \bibinfo {pages} {170403} (\bibinfo
  {year} {2007})}\BibitemShut {NoStop}%
\bibitem [{\citenamefont {Young}\ and\ \citenamefont
  {Frederikse}(1973)}]{youngCompilationStaticDielectric1973}%
  \BibitemOpen
  \bibfield  {author} {\bibinfo {author} {\bibfnamefont {K.~F.}\ \bibnamefont
  {Young}}\ and\ \bibinfo {author} {\bibfnamefont {H.~P.~R.}\ \bibnamefont
  {Frederikse}},\ }\bibfield  {title} {\enquote {\bibinfo {title} {Compilation
  of the {{Static Dielectric Constant}} of {{Inorganic Solids}}},}\ }\href
  {\doibase 10.1063/1.3253121} {\bibfield  {journal} {\bibinfo  {journal}
  {Journal of Physical and Chemical Reference Data}\ }\textbf {\bibinfo
  {volume} {2}},\ \bibinfo {pages} {313--410} (\bibinfo {year}
  {1973})}\BibitemShut {NoStop}%
\bibitem [{\citenamefont {Carlesso}\ and\ \citenamefont
  {Bassi}(2016)}]{carlessoDecoherenceDueGravitational2016}%
  \BibitemOpen
  \bibfield  {author} {\bibinfo {author} {\bibfnamefont {Matteo}\ \bibnamefont
  {Carlesso}}\ and\ \bibinfo {author} {\bibfnamefont {Angelo}\ \bibnamefont
  {Bassi}},\ }\bibfield  {title} {\enquote {\bibinfo {title} {Decoherence due
  to gravitational time dilation: {{Analysis}} of competing decoherence
  effects},}\ }\href {\doibase 10.1016/j.physleta.2016.05.034} {\bibfield
  {journal} {\bibinfo  {journal} {Physics Letters A}\ }\textbf {\bibinfo
  {volume} {380}},\ \bibinfo {pages} {2354--2358} (\bibinfo {year}
  {2016})}\BibitemShut {NoStop}%
\bibitem [{\citenamefont {Nguyen}\ and\ \citenamefont
  {Bernards}(2020)}]{nguyenEntanglementDynamicsTwo2020}%
  \BibitemOpen
  \bibfield  {author} {\bibinfo {author} {\bibfnamefont {H.~Chau}\ \bibnamefont
  {Nguyen}}\ and\ \bibinfo {author} {\bibfnamefont {Fabian}\ \bibnamefont
  {Bernards}},\ }\bibfield  {title} {\enquote {\bibinfo {title} {Entanglement
  dynamics of two mesoscopic objects with gravitational interaction},}\ }\href
  {\doibase 10.1140/epjd/e2020-10077-8} {\bibfield  {journal} {\bibinfo
  {journal} {The European Physical Journal D}\ }\textbf {\bibinfo {volume}
  {74}},\ \bibinfo {pages} {69} (\bibinfo {year} {2020})}\BibitemShut {NoStop}%
\end{thebibliography}
\end{document}